\documentclass{llncs}

\usepackage{ifpdf}
 \ifpdf
\pdfoutput=1
\usepackage[pdftex]{graphicx}
\newcommand{\pix}{jpg}
 \else
\usepackage[dvips]{graphicx}
\newcommand{\pix}{bmp}
 \fi

\usepackage{url}
\usepackage{amsmath}
\usepackage{amssymb}

\def\url#1{{\tt{#1}}}

\renewcommand{\thefigure}{\arabic{subsection}}
\renewcommand{\thefigure}{\arabic{figure}}

\begin{document}
\spnewtheorem{coro}[subsubsection]{Corollary}{\bfseries}{\rmfamily}
\spnewtheorem{defi}[subsection]{Definition}{\bfseries}{\rmfamily}
\spnewtheorem{lem}[subsection]{Lemma}{\bfseries}{\rmfamily}

\def\mytitl{Constructive Non-Linear Polynomial Cryptanalysis~}
\def\mytitm{of a Historical Block Cipher}
\title{\mytitl \mytitm}
%

\author{
Nicolas T. Courtois
\and
Marios Georgiou
}
\authorrunning{N. T. Courtois}
\tocauthor{\protect{}}

\institute{
University College London, Gower Street, London, UK
}

\maketitle

\vskip-10pt
\vskip-10pt
\begin{abstract}
One of the major open problems in symmetric cryptanalysis
is to discover new specific types of invariant properties which
can hold for a larger number of rounds of a block cipher.
We have Generalised Linear Cryptanalysis (GLC)
and Partitioning Cryptanalysis (PC).
Due to double-exponential combinatorial explosion of
the number of possible invariant properties systematic exploration is not possible 
and extremely few positive working examples of GLC are known.
Our answer is to work with polynomial algebraic invariants which makes partitions more intelligible.
We have developed a constructive algebraic approach which is about
making sure that a certain combination of polynomial equations is zero.
We work with an old block cipher from 1980s which
has particularly large hardware complexity compared to modern ciphers e.g. AES.
However all this complexity is not that useful if we are able to construct
powerful non-linear invariants which work for any number of rounds.
A key feature of our invariant attacks is that
we are able to completely eliminate numerous state and key bits.
We also construct invariants for the (presumably stronger) KT1 keys.
Some of these 
lead to powerful ciphertext-only correlation attacks.
\vskip-3pt
\vskip-3pt
\end{abstract}


\newpage

\vskip-6pt
\vskip-6pt
\section{Introduction, Non-Linear Cryptanalysis}
\vskip-6pt

The concept of cryptanalysis with non-linear polynomials
a.k.a. Generalized Linear Cryptanalysis (GLC)
was introduced by Harpes, Kramer, and Massey at Eurocrypt'95, cf. \cite{GenLinear1}.
A key question is the existence of round-invariant I/O sums: 
when a value of a certain polynomial is preserved
after 1 round.
Many researchers have in the past failed to find any such properties,
cf. for example Knudsen and Robshaw at Eurocrypt'96 cf. \cite{GenLinear2} 
and there are extremely few positive results on this topic,
cf. recent result 
\cite{TodoNL18}. 
A paper at Crypto 2004 cf. \cite{BLC} provides the following hint: 
non-linear polynomial I/O sums 
can eventually be made to work if we consider that the choice of monomials 
will strongly depend on the structure and internal wiring 
of the cipher.
Bi-Linear and Multi-Linear cryptanalysis were subsequently introduced \cite{BLC,invglc} 
in order to work with Feistel ciphers with two and several branches specifically.
In this paper we revisit the question of non-linear cryptanalysis and give it a fresh start.
Rather than building from scratch some highly contrived
new ciphers
\cite{BannierPartBack,invglc,WhiteningParadox} 
we are going to {\bf construct} a more substantial variety of round invariant properties,
on demand,
for a given (more natural) real-life block cipher setting.
Our work is rather disjoint w.r.t. recent high-profile results in \cite{TodoNL18}
dealing primarily with SPN ciphers, we focus more on 
Feistel ciphers.
Moreover unlike in \cite{TodoNL18} we focus on invariants which work for 100 $\%$
of the keys and we focus on stronger invariants which hold with probability equal to 1
(accordingly the key schedule is less relevant here).

\vskip-6pt
\vskip-6pt
\subsection{Combinatorial or Algebraic, Partitioning Cryptanalysis}
\vskip-3pt

A classical open problem in cryptanalysis is discovery
of invariant or semi-invariant properties
of complex type, cf. recent paper \cite{TodoNL18}.
The space of solutions is double exponential and many authors
stress that systematic exploration is rather
impossible \cite{BeiCantResNL}.
There are two major approaches to our problem: combinatorial and algebraic.
A naive combinatorial approach would be to try to modify the connections of the cipher
and test (more or less at random) for a desired property.
A better 
approach is to make more constants variable
and consider constraint satisfaction problems with formal coding and SAT solvers.
We propose a more meaningful and compact ``algebraic'' approach
with 
polynomial equations.
A major novelty here is to consider that Boolean functions are also {\bf unknowns}
which can be determined rather than fixed and
they could be weak maybe {\bf only} w.r.t one very specific cipher setup.
Our multivariate polynomial coding is compact and allows precise
understanding when and why a cipher could be weak 
and many very precise results (up to ``if and only if type''). 

A well known very general (combinatorial) approach
considers arbitrary subsets of 
binary vector spaces,
and it is called Partitioning Cryptanalysis (PC),
cf. \cite{BannierPartBack,
HarpMassThm,JakobsenPartitioning}.
A more algebraic approach is to consider only specific forms of partitions,
mainly those defined by the value (0 or 1) of a single Boolean polynomial.
This is of course less general, BUT it leads to a more {\bf effective} approach to the problem,
effective in the sense that we expect that properties are described, discovered and studied with
the tools of algebra. In particular new very interesting questions can be asked,
and we expect properties to be computed or derived rather than to happen by some incredible coincidence.
In other terms we expect the algebraic approach to be more illuminating about what actually happens
and also easier to study.
More importantly in this paper we put forward an 
{\bf algebraic discovery} approach of a new sort.
Our problem will be coded with a surprisingly simple single equation
of a limited degree which we will call FE.
Solving such equation(s), there will be up to 8 such equations as we will see,
{\bf guarantees} that we obtain a Boolean function
and the polynomial invariant ${\cal P}$ which makes a block cipher weak.
Moreover when this equation FE
reduces to zero, our invariant becomes stronger
in the sense of holding for a larger space of Boolean functions,
cf. Section \ref{FEReductionToZeroExample2b} and Section
\ref{SimpleInvP20Cycle9BiasedFEreducedto0}.
Solving FE also avoids exploring the vast
space of possible Boolean functions
(this again would be a combinatorial approach).
Specific examples will be constructed based on a
highly complex historical
block cipher.
We construct several examples where the set of solutions is not empty
which demonstrates that
our attack actually works.

\vskip-9pt
\vskip-9pt
\subsection{Weak vs. Strong Keys and Ciphers}
\vskip-5pt

The question of weak keys has not received sufficient attention in cryptography research.
The single key attack is certainly not a correct way to evaluate a security of a cipher.
Almost every cipher in the real life is used with multiple random keys, and the attacker only
needs to break some fraction of the keys.
There are numerous constructions of weak ciphers in cryptographic literature,
cf. for example work related to the AES S-box \cite{invglc,WhiteningParadox},
and very recent constructions in 
\cite{BannierPartBack,FiliolNotVuln}.
A serious theory is nowadays being developed around what is possible or not to
achieve in partitioning and invariant attacks,
with important notions of strongly proper transformations and anti-invariant resistance,
cf. \cite{BeiCantResNL,FiliolNotVuln}.
There are two major types of invariants in recent research:
linear or sub-space invariants, 
and proper non-linear invariants.
Several authors \cite{BeiCantResNL,FiliolNotVuln} including
this paper study both.
Almost all research in this area revolves around
the fundamental notion of Partitioning\footnote{
In fact what we do in this paper is ALSO partitioning cryptanalysis, except
that our partitions are characterized by a [single] multivariate polynomial.}
 Cryptanalysis (PC) cf.
\cite{BannierPartBack,
HarpMassThm}
and 
PC can be linear or non-linear and generalizes both LC and GLC \cite{
HarpMassThm}.
These works are closely related and also to
the study of the groups generated by various cipher
transformations
\cite{T-310An80,
KennyImprimitive,AESWernsdorf,
invglc,WhiteningParadox}.
In this paper we contend that 
the partitioning approach is 
{\bf too general}. 
It helps to 
establish some impossibility results \cite{BeiCantResNL,FiliolNotVuln},
yet
it obscures any possibility results.

\vskip-8pt
\vskip-8pt
\subsubsection{What's New?}
\vskip-5pt
We can discover some invariant properties but do we understand their nature?
Can we manipulate the properties efficiently and compress them (represent in a compact way)?
Can we discover properties with some effective computational methods and see how various
constraints will imply their existence or not?
Can we show that some ciphers are going to be secure against such attacks?
Can we construct weak keys secure w.r.t. simpler (e.g. linear) invariants
and 
previously known attacks?
Yes we can,
and our polynomial algebraic discovery approach
is what makes all these possible.

\vskip-8pt
\vskip-8pt
\subsubsection{Mathematical Theory of Invariants.}
There exists an extensive theory of multivariate polynomial algebraic invariants 
going back to 1845 \cite{
CrillyInvariantsHist}. 
However mathematicians have studied primarily invariants w.r.t. linear transformations(!).
Moreover the classical invariant theory has very rarely considered invariants with more than 5 variables
and in finite fields of small size. 
In our work we study invariants w.r.t {\bf non-linear} transformations (!!!) and with up to 36 variables over $GF(2)$.
A well-known polynomial invariant with applications in symmetric cryptography is the cross-ratio, 
studied in Sect. 4 in \cite{invglc} and
in the ``whitening paradox''\footnote{The ``paradox'' is a proof of concept that a group-theoretic claims
in cryptography \cite{T-310An80,
AESWernsdorf} 
can be highly misleading
and can lead to ciphers which are insecure \cite{WhiteningParadox}.} cf. \cite{invglc,WhiteningParadox}.
%


\vskip-9pt
\vskip-9pt
\section{Notation and Methodology}
\vskip-5pt

In this paper we are going to work with one specific block cipher in order to show
that non-linear cryptanalysis can be made to work.
We do not provide a full description of an encryption system
(how it is initialized and used etc).
We just concentrate on how one block cipher round operates
and how it translates into relatively simple Boolean polynomials.
Quite importantly, we consider, which is rarely done in cryptanalysis,
that the Boolean function is an unknown, yet to be determined.
We will denote this function by a special variable $Z$.
We will then postulate that $Z$ may satisfy a certain algebraic equation
[with additional variables] and then this equation will be solved
in order to determine $Z$.

In order to have notations, which are as compact as possible,
in this paper the sign + will denote addition modulo 2, frequently we will omit the sign * in products
and will frequently use short or single letter variable names.
In general in this paper we will use small letters $a-z$,
and $x_{1},x_{36}$ or $e_1$ for various binary variables $\in\{0,1\}$.
Certain capital letters $S1,S2,K,L,F,Z$ will be used to represent some very ``special''
sorts of variables which are placeholders for something more complex.
In particular the capital letter $Z$
is a placeholder for
substitution of the following kind

\vskip-4pt
\vskip-4pt
$$
Z(e_1,e_2,e_3,e_4,e_5,e_6)
$$
\vskip-3pt

where $e_1\ldots e_6$ will be some 6 of the other variables.
In practice, the $e_i$ will represent a specific subset of variables
of type $a$-$z$, or other such as $L$,
therefore at the end, our substitution will actually look like:

\vskip-7pt
\vskip-7pt
$$
Z \leftarrow Z00+Z01*L+Z02*c+Z03*Lc+\ldots +Z62*cklfh+Z63*Lcklfh
$$
\vskip-4pt

Other capital letters will be used to signify some bits which are also
unknown which will be bits of the secret key used in a given round
and such bits are in our work called by letters 
$S1,S2$, where S2 will be sometimes renamed $L$
and $S1$ can be also called $K$.
We also use a capital letter to represent some bits which depend on the IV,
or a round-dependent variable constants.
This sort of variable is typically known to the attacker and
will be denoted by the letter $F$.
Our notations would typically omit to specify in which round of encryption
these bits are taken, as most of our work is about constructing {\bf one round} invariants
(which however do extend to an arbitrarily large numbers of rounds).
We consider in general that each round of encryption will be identical except that
they can differ only in some ``public'' bits called $F$ (and known to the attacker)
and some ``secret'' bits called 
$S1,S2$ or $K,L$ and unknown to the attacker.
These bits will be different in different rounds.
This framework covers most block ciphers ever made  
except that some ciphers will have more ``secret'' bits in one round.


\vskip-8pt
\vskip-8pt
\subsection{Polynomial Invariants}
\label{FirstGlimpseOnPSymmetries}
\vskip-2pt

We are looking for arbitrary polynomial invariants coded as formal polynomials
and the structure of the polynomial is provided or guessed by the attacker depending
on 
``heuristics'' 
based on the high-level structure of the cipher
following the ideas of
\cite{BLC}.
%
This frequently, but not\footnote{
Some of our polynomials 
are very far from being symmetric, e.g. in Section \ref{Example317NotSymmetric}.
we have ${\cal P}$ such that ${\cal P}-fg$ is a symmetric homogenous polynomial,
in another example ${\cal P}$ has only
2 monomials of degree 2, cf. Section \ref{SimpleInvP20TwoProducts}.
}
at all in general, 
will lead to combinations of symmetric polynomials like:

\vskip-8pt
\vskip-8pt
$$
{\cal P}(a,b,\ldots) = P51*(e+f+g+h) + P42*(abc+abd+acd+bcd) + \ldots
$$
\vskip-2pt

\noindent
We will avoid completely general polynomials which would be too costly to consider and process efficiently.
In addition we will attempt to reduce (e.g. through a row-echelon process studied later in Section \ref{ExampleF0F1andL0L01withZ1andZ4Spaces})
the number of variables such as P51 determined at a later ``algebraic solving'' stage of our attack.
%
%
%
In this paper we
try to strike a balance between a completely general
constructive approach applicable to any block cipher,
and constructing simple examples 
dictated by the high-level structure
of one specific Feistel cipher.
%
%
%
%
%
In theory, our approach is totally general
as every function could be written as a polynomial over a finite field,
therefore in a very broad (and naive) sense Partitioning Cryptanalysis (PC) and
Generalized Linear Cryptanalysis are equivalent.
In practice, we will be working with sets characterized by one polynomial at a time,
polynomials are specific type, 
and our computational ability to solve the equations is limited.


\newpage

\subsection{Our Specific Cipher}
\vskip-5pt

In this paper we are going however to work primarily on a particularly
complex Feistel cipher on 4 branches known as T-310.
T-310 was the principal encryption algorithm used to protect various state communication lines
in Eastern Germany, and until 1989 some 3,800 cipher machines were in active service \cite{FeistCommunist,T-310,MasterPaperT310,T-310An80}.
The block size is 36 bits and the key has 240 bits.
We are going to explore the 
space of invariants on 36 bits and study Boolean functions on $6$ bits,
some of which may lead to specific invariant attacks.
There are $2^{2^{6}}=2^{64}$ Boolean functions on 6 bits
and an incredibly large number $2^{2^{36}}$ of possible invariants.

\vskip-7pt
\vskip-7pt
\begin{figure}
\vskip-7pt
\begin{center}
\includegraphics*[width=5.1in,height=3.2in]{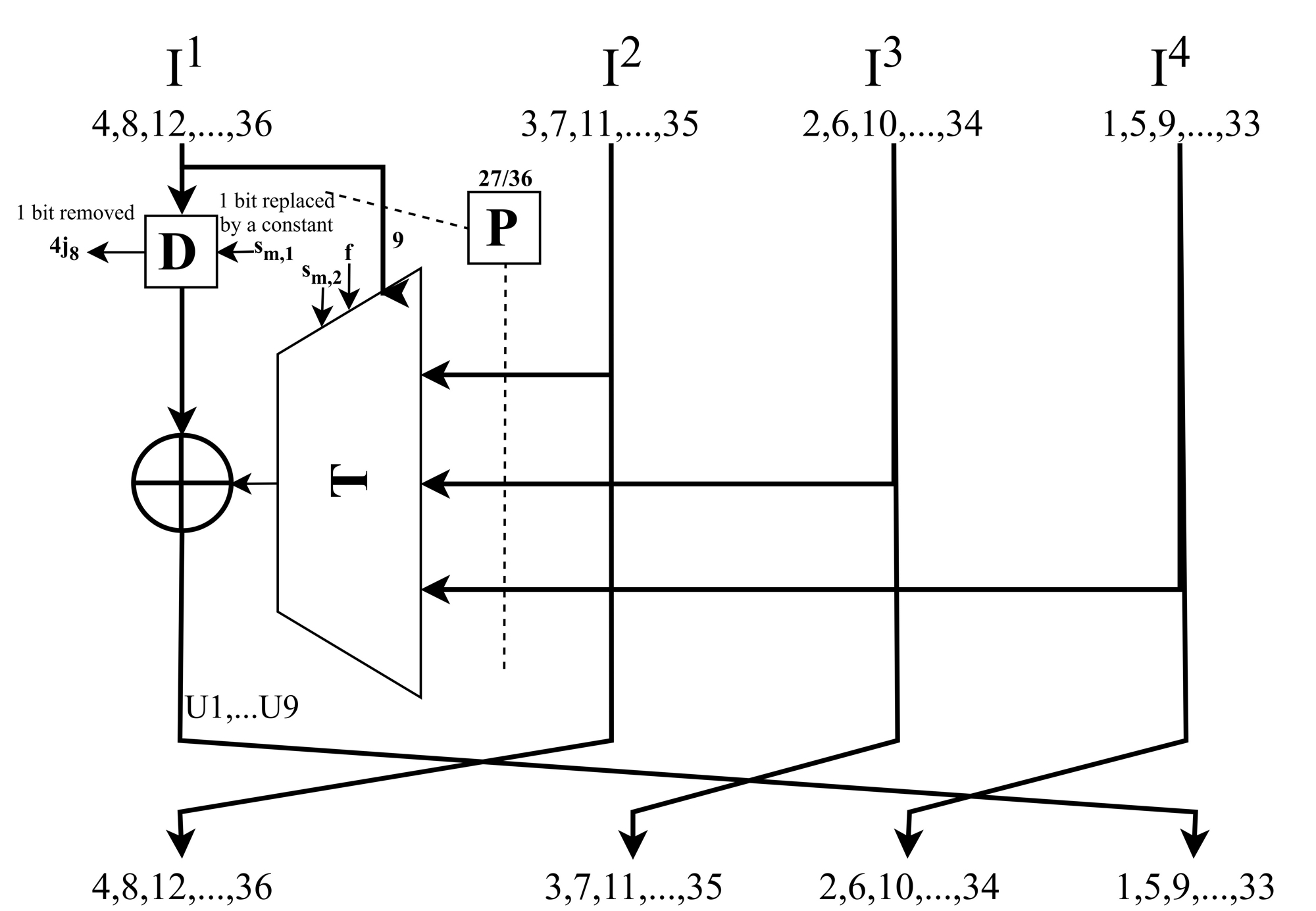}
\end{center}
\vskip-7pt
\vskip-7pt
\caption{T-310: a peculiar sort of Compressing Unbalanced Feistel scheme cf. \cite{UnbContractPata,MasterPaperT310}.
}
\label{FigContracting310KT1}
\vskip-7pt
\vskip-7pt
\end{figure}
\vskip-7pt
\vskip-7pt

\vskip-8pt
\vskip-8pt
\subsection{Constructive Approach Given the Cipher Wiring}
\vskip-6pt

We decided to execute our task on a cipher which offers great
{\bf flexibility} in the choice of the internal wiring,
so that we can possibly make such adjustments if we do not find a property we are looking for.
Most ciphers such as DES or AES also have this sort of flexibility in the choice of P-boxes,
arbitrary invertible matrices inside the S-box, inside the mixing layers, etc.
Here we work with the T-310 cipher from 1980s where such changes are officially allowed and are
officially specified by the designers of the cipher.
Here if we find a weak setup, and we will,
it can be directly implemented with original historical hardware.
Our work is at the antipodes compared to \cite{invglc,WhiteningParadox}
where the ciphers are really very special and have very strong 
high-level structure.
Our approach is really the opposite:
we start from any given cipher spec in forms of ANFs for one round
and we generate complex invariant properties essentially on demand.
%
Our approach is applicable to more or less
any block cipher which contains non-linear components,
and also to most hash functions and stream ciphers
based on a core block cipher.




In the picture below we show the internal structure of T-310,
one of the most important block ciphers of the Cold War, 
massively used 
to encrypt all sorts of state communications, cf. \cite{T-310}.
T-310 
is one of the most ``paranoid'' cipher designs we have ever seen.
The cipher is iterated hundreds of times per one bit actually encrypted.
The hardware complexity of T-310 is hundreds of times bigger than AES or 3DES,
cf. \cite{LCKT1ucry18,FeistCommunist}.
Does it make this cipher very secure?
Not quite, if we can 
construct
algebraic invariants which work for any number of rounds.


%

\vskip-8pt
\vskip-8pt
\begin{figure}
\vskip-5pt
\vskip-5pt
\hskip-10pt
\hskip-10pt
\begin{center}
\hskip-10pt
\hskip-10pt
\vskip-6pt
\vskip-6pt
\includegraphics*[width=5.1in,height=2.8in]{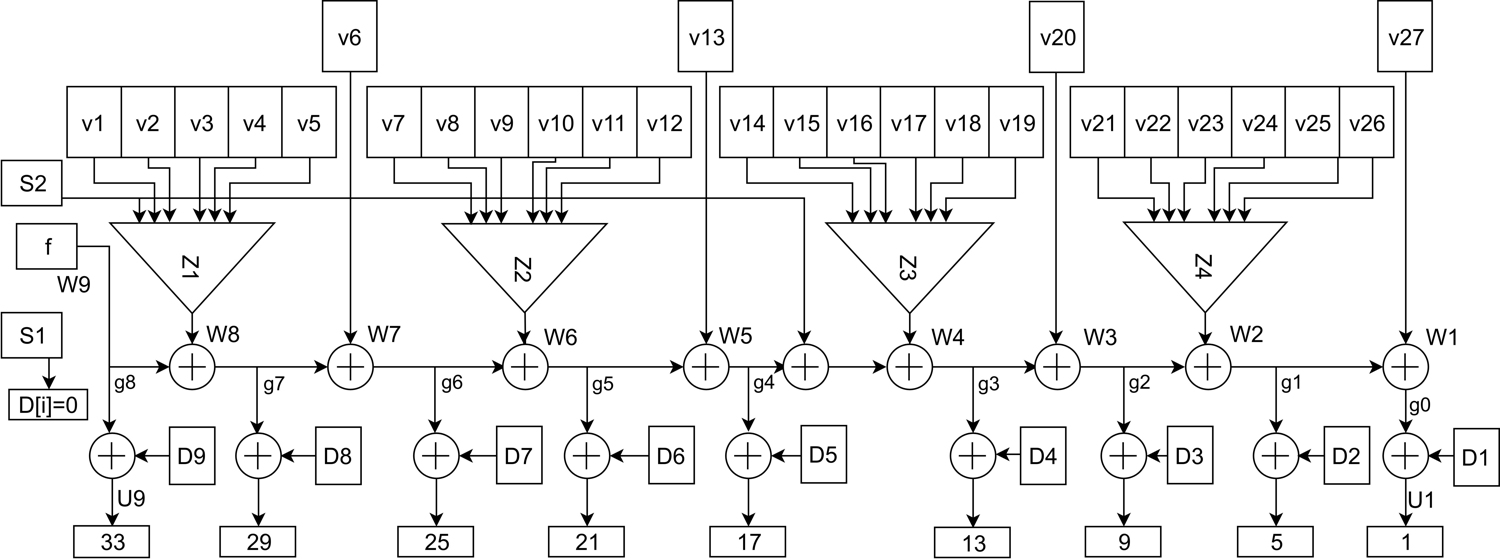}
\end{center}
\vskip-9pt
\vskip-9pt
\caption{The internal structure of one round of T-310 block cipher.
}
\label{FigComplicationUnit6basic}
\vskip-6pt
\vskip-6pt
\end{figure}
\vskip-2pt
\vskip-2pt

The cipher operates on 36-bit blocks and the state bits are numbered 1-36.
The bit numbering in this compressing unbalanced Feistel cipher with 4 branches
is such, cf. Fig \ref{FigContracting310KT1}, that bits $1,5,9\ldots 33$ are those freshly created by this round,
while ALL the input bits the numbers of which are NOT multiples of 4 are shifted by 1 position, i.e.
bit 1 becomes 2 in the next round, and bit 35 becomes 36.
Few things remain unspecified on our picture:
which bits and in which order are connected to D1-D9 and v1-v27.
In T-310 this specification is called an
LZS or {\em Langzeitschl\"{u}ssel}
which means a long-term key and which is distinct than the short-term key on 240 bits.
We simply need to specify
two functions\footnote{
Which are both assumed to be injective and $D(i)$ are always multiples of $4$,
this in order to avoid many degenerate cases and trivial attacks cf. \cite{MasterPaperT310}.}
$D: \{1\ldots 9\} \to \{0\ldots 36\}$, $P:\{1\ldots 27\}\to \{1\ldots 36\}$.
For example $D(5)=36$ will mean that input bit 36 is connected
to the wire D5 on our picture, and $P(1)=25$ will mean that input 25 is connected
as v1 or the 1st input of $Z1$.
Finally the internal wiring LZS uses a special convention where the bit S1 is used instead of one
of the $Di$ by specifying
 that $D(i)=0$.
%
%
Overall one round can be described as 36 Boolean polynomials out of which only 9 are non-trivial.
Let $x_{1},\ldots ,x_{36}$ be the inputs and
let $y_{1},\ldots ,y_{36}$ be the outputs.

\vskip-9pt
\vskip-9pt
\begin{align*}
y_{33}  &= F + x_{D(9)}  & \\
& Z1 \hspace{-.28em}\stackrel{def}{=}\hspace{-.28em}Z(S2,x_{P(1)},\hspace{-.20em}\ldots,\hspace{-.20em}x_{P(5))})\hspace{-.55em}\hspace{-.70em}\\
y_{29}  &=  F  + Z1 + x_{D(8)} & \\
y_{25}  &=  F  + Z1 + x_{P(6)}+x_{D(7)} & ~~~~~~~~~~~~~~~~~~~~~~~~~~~ &\\
& Z2 \stackrel{def}{=}Z(x_{P(7)},\ldots, x_{P(12)})\\
y_{21}  &=  F  + Z1 + x_{P(6)}+Z2+ & x_{D(6)} ~~~~~~~~~~~~~~~~~~~~~~~~~~~~& \\
y_{17}  &=  F  + Z1 + x_{P(6)}+Z2+ &  x_{P(13)} + x_{D(5)} ~~~~~~~~~~~~~~~~& \\
& Z3 \stackrel{def}{=}Z(x_{P(14)},\ldots, x_{P(19)})\\
y_{13}  &=  F  + Z1 + x_{P(6)}+Z2+ & x_{P(13)} + S2 + Z3 + x_{D(4)} ~& \\
y_{9}  &=  F  + Z1 + x_{P(6)}+Z2+ & x_{P(13)} + S2 + Z3 + x_{P(20)} &+x_{D(3)}~~~~~~~~~~~~~~~~~ \\
& Z4 \stackrel{def}{=}Z(x_{P(21)},\ldots, x_{P(26)})\\
y_{5}  &=  F  + Z1 + x_{P(6)}+Z2+ & x_{P(13)} + S2 + Z3 + x_{P(20)} &\hspace{-.15em}+\hspace{-.15em}Z4 \hspace{-.15em}+\hspace{-.15em} x_{D(2)}\hspace{-.15em}~~~~~~~~~~~~  \\
y_{1}  &=  F  + Z1 + x_{P(6)}+Z2+ & x_{P(13)} + S2 + Z3 + x_{P(20)} &\hspace{-.15em}+\hspace{-.15em}Z4 \hspace{-.15em}+\hspace{-.15em} x_{P(27)} \hspace{-.15em}+\hspace{-.15em} x_{D(1)} \\
 & x_{0} \stackrel{def}{=}S1 & \\
y_{i+1}&=x_{i} \mbox{~for all other~} i\ne 4k & (\mbox{~with~} 1\leq i\leq 36)~~~~~~~~~~\\
\end{align*}
\vskip-7pt
\vskip-7pt

\noindent
Here 
$F$ is a public bit derived from an IV transmitted in the cleartext,
S1 and S2 are bits of the secret key on 
240 bits. S1 and S2 are repeated every 120 steps,
and the key scheduling is weak,
cf. our correlation attack in 
Section \ref{CiphertextOnlyAttacks}.


%
We will now migrate towards more convenient and shorter notations.
We emphasise the fact that the variables $x_i$ and $y_i$ are treated ``alike''
and we will call them by lowercase letters a-z backwards starting from
$x_{36}$ till $x_{11}$ and $y_{36}$ till $y_{11}$. Then we use
capital letters $M$-$V$ (avoiding some letters used elsewhere). 
For example $a=x_{36}$ and $t=x_{17}$, and $M=x_{10}$ and $V=x_{1}$.
The reader may be surprised to see that in our cryptanalysis efforts we
attempt to map $x_i$ to $y_i$ (they are denoted by the same letter)
and this for ALL the 36 variables simultaneously and this after exactly 1 round.
This even though for almost all variables $x_i=y_{i-1}$ after one round.
This is {\bf not impossible}, not even in Linear Cryptanalysis (LC).
For example if
$a \stackrel{def}{=}x_{36}$ in input expressions, and also used to denote $y_{36}$ for output expressions
$b \stackrel{def}{=}y_{35}$ and also used to denote $y_{35}$,
it is unthinkable that say $a$ is an invariant property if we replace $a\leftarrow b$
in 
the next round.
However 
$a+b+c+d$ absolutely CAN be an invariant 
for one round (!).
This, only IF our cipher had ANY invariant linear property
true with probability 1 for 1 round which
is possible but not so common.
A recent paper shows that this
can occur for about $3\%$
of the so called 
officially approved KT1 keys
cf. \cite{LCKT1ucry18}.
In our paper we will be of course looking at
what happens for the remaining {\bf stronger} 97 $\%$ of LZS keys
and at more general {\bf non-linear} polynomial expressions such as
say $ef+fg+eh+gh$. 
All examples we give in this paper are real-life examples.
%


\vskip-7pt
\vskip-7pt
\subsubsection{The Result After Renaming.}
\label{ExactSubstitutions}
Overall, depending on the exact values of $D(i)$ and $P(j)$,
we can rewrite the beginning of our equation system as follows.
The variables on the left hand side will be output variables after 1 round,
and on the right hand side, we have ANF or polynomials in the input variables.


\vskip-9pt
\vskip-9pt
\begin{align*}
a  &\leftarrow  b & \\
b  &\leftarrow  c & \\
c  &\leftarrow  d & \\
d  &\leftarrow  F + i & \\
& [\ldots] & \\
h  &\leftarrow  F + Z + e & \\
& [\ldots] & \\
&Z1 \leftarrow Z(L,j,h,f,p,d))\\
z  &\leftarrow M & \\
M  &\leftarrow N & \\
N  &\leftarrow F + Z + r + Y + m+ L + X +x +k & \\
& [\ldots] & \\
& Z4 \leftarrow Z(a,g,c,z,U,i)\\
R &\leftarrow F + Z + r + Y + m+ L + X + W + x&\\
& [\ldots] & \\
V &\leftarrow F + Z + r + Y + m+ L + X + W + x + w&\\
\end{align*}
\vskip-7pt
\vskip-7pt
\label{SubsEqsFor317}
\vskip-7pt

\noindent
These expressions should be viewed as a sequence of substitutions
where a variable is replaced by a polynomial algebraic expression.
%
%
In order to have shorter expressions to manipulate 
we replaced here $Z1-Z4$ by shorter abbreviations $Z,Y,X,W$ respectively.
We also replaced S2 by a single letter $L$ (used at 2 places).
The other key bits $S1$ will only be used at one place if some $D(i)=0$.
%
%
An observation is that the algebraic degree
of all our ANF expressions is constant and at most 6.
%
%
Now the only thing which remains to be done is to find a polynomial
expression ${\cal P}$ say

\vskip-5pt
\vskip-5pt
$$
{\cal P}(a,b,c,d,e,f,g,h,\ldots) =
abcdijkl+efg+efh+egh+fgh
$$
\vskip-1pt

using any number between 1 and 36 variables
such that if we substitute in ${\cal P}$  all the variables by the substitutions defined
above in Section \ref{ExactSubstitutions},
we would get exactly the same polynomial expression ${\cal P}$ .

%

\newpage

\vskip-6pt
\vskip-6pt
\section{The Fundamental Equation}
\vskip-1pt


We want to {\bf find} a polynomial
expression ${\cal P}$ using any number between 1 and 36 variables such that
it is an invariant after the substitutions of Section \ref{ExactSubstitutions}.
%
For example if the polynomial ${\cal P}$ is fixed\footnote{
One simple method is to select a specific short symmetric polynomial which have already been seen to work.
In practice ${\cal P}$ is not completely fixed and
rather selected to belong to a certain (reduced) space of multivariate polynomials.},
and also in other cases,
the attacker will write ONE SINGLE (or more) algebraic equation which he is going to solve
to determine the unknown Boolean function $Z$
(and ${\cal P}$) if a solution exists.

\begin{defi}[Compact Uni/Quadri-variate FE]
\label{defiCompactFE}
Our ``Fundamental Equation (FE)'' to solve is simply a substitution like:
\vskip-3pt
\vskip-3pt
$$
{\cal P}(Inputs) =
{\cal P}(Outputs)
$$
\vskip-3pt

or more precisely

\vskip-5pt
\vskip-5pt
$$
{\cal P}(a,b,c,d,e,f,g,h,\ldots) =
{\cal P}(b,c,d,F+i,f,g,h,F+Z1+e,\ldots)
$$
\vskip-3pt

\end{defi}
\vskip-3pt
\vskip-3pt

where again $Z1-Z4$ are replaced by $Z,Y,X,W$.
In the next step, $Z$ will be represented by an Algebraic Normal Form (ANF)
with 64 binary variables which are the coefficients of the ANF of $Z$,
and there will be several equations,
and four {\bf instances} $Z,Y,X,W$ of the same $Z$:

\begin{defi}[A Multivariate FE]
\label{defiRewriteFEZ00}
At this step 
we will rewrite FE as follows. We will replace Z1 by:

\vskip-6pt
\vskip-6pt
$$
Z \leftarrow Z00+Z01*L+Z02*j+Z03*Lj+\ldots +Z62*jhfpd+Z63*Ljhfpd
$$
\vskip-1pt

Likewise we will also replace $Z2$:
\vskip-6pt
\vskip-6pt
$$
Y \leftarrow Z00+Z01*k+Z02*l+Z03*kl+\ldots +Z62*loent+Z63*kloent
$$
\vskip-1pt
\noindent
and likewise for $X=Z3$ and $W=Z4$ and the coefficients $Z00\ldots Z63$
will be the same inside $Z1-Z4$, however the subsets of 6 variables
chosen out of 36 will be different in $Z1-Z4$.
Moreover, some coefficients of ${\cal P}$ may also be variable.
\end{defi}


%
Initially, we can select ${\cal P}$ as an arbitrary fixed polynomial, with degree say between 2 and 26.
Then if we cannot find a solution,
we will enlarge the space of solutions but making more
or all coefficients of ${\cal P}$ variable.
In all cases, all we need to do is to solve the equation above for $Z$,
plus a variable amount of extra variables.
This formal algebraic approach, if it has a solution,
still called $Z$
for simplicity, 
or $({\cal P},Z)$
will {\bf guarantee} that our invariant ${\cal P}$ holds for 1 round.


%
%
%
%

The process is as we can see EXTREMELY SIMPLE:
we assume that a certain equation holds for $Z$ and we solve it for $Z$
which is 64 binary unknowns for the ANF coefficients,
(for example using a SAT solver).
Our experience seems to show that this problem will rarely be actually computationally really hard.
This depends on many factors such as size, degree and shape of ${\cal P}$.
Moreover the dimension of the space of ${\cal P}$ can be reduced by
specific ad-hoc requirements decided by the attacker,
cf. Section \ref{ExampleF0F1andL0L01withZ1andZ4Spaces}.

%
%
%
%
%

\subsection{The Solvability Problem}
\vskip-6pt

A major problem is now {\bf the existence of solutions}.
Does this equation FE have a solution? Or, does it have solutions
which would satisfy some additional requirements required to design
any sort of meaningful cryptanalytic attack?
%
There are, as the reader may guess, countless cases
where this problem has {\bf no solution} whatsoever.
%
In many interesting cases this equation will be unusually simple
and sometimes it will disappear totally,
cf. Sections \ref{FEReductionToZeroExample2b} and \ref{SimpleInvP20Cycle9BiasedFEreducedto0}.


\vskip-6pt
\vskip-6pt
\subsection{What is New}
\vskip-6pt

A previous paper on this topic \cite{BackdTut} emphasises
invariants of low degree which are irreducible in cases without linear attacks
and when the Boolean function and the long-term keys quite special.
In this paper we are looking for better attacks.
We show that invariants of higher degrees are more powerful and
lead to internal correlation attacks which can also work with
KT1 keys and for any Boolean function.

\vskip-6pt
\vskip-6pt
\section{A Toy Example which Does Not Work Well}
\vskip-6pt

Our approach is to construct toy ciphers fully compliant with the spec of T-310 cipher
and which are weak w.r.t. non-linear cryptanalysis.
In general countless interesting invariants exist for this cipher.
These examples were constructed rather than found,
by combination of paper and pencil work on FE with some heuristics and ad-hoc assumptions,
with some computer simulations in order to satisfy some additional criteria,
mainly resistance to other known attacks,
and the ability to eliminate many state and key bits
so that they can be eventually used to cryptanalyse our cipher.
We start with some toy examples chosen for their elegance and simplicity which do not yet work very well.

\vskip-7pt
\vskip-7pt
\subsection{A Toy Example with ${\cal P}$ of Degree 3 and 4 Variables and $F=0$}
\label{Example317NotSymmetric}
\vskip-6pt

We start with the following very simple non-linear round invariant property ${\cal P}$
which is particularly simple and uses only 4 variables:

\vskip-4pt
\vskip-4pt
$$
{\cal P}=efg+efh+egh+fgh+fg
$$
\vskip-2pt

Our polynomial is such that ${\cal P}-fg$ is a symmetric homogenous polynomial of degree 3 in 4 variables only.
However it is important to note that overall ${\cal P}$ is {\bf NOT} a symmetric polynomial.
and it is also irreducible\footnote{We refer to later Section \ref{CiphertextOnlyAttacks}
and Appendix \ref{SimpleInvP20Cycle9BiasedFEreducedto0App} to see why this matters.}.
Consider the following long term key setup:

\vskip-5pt
\vskip-5pt
\begin{verbatim}
317: P=27,29,31,21,33,19,26,25,22,32,23,17,24,16,18,9,5,
10,35,13,36,30,34,11,2,28,14 D=17,25,26,35,18,34,30,32,28
\end{verbatim}
\vskip-2pt

\noindent
What is the solution $Z$?
All we need to do is to write our Fundamental Equation (FE)
and substitute four variables using the
ANF round equations in page \pageref{SubsEqsFor317}.

\vskip-6pt
\vskip-6pt
$$
{\cal P}(a,b,c,d,e,f,g,h,\ldots) =
{\cal P}(b,c,d,F+i,f,g,h,F+Z1+e,\ldots)
$$
\vskip-2pt

On the right hand side we replace $a\leftarrow b$, etc,
up to $h  \leftarrow  F+Z1+e$.
%
%

Our Fundamental Equation (FE) becomes:

\vskip-4pt
\vskip-3pt
$$
F(fg+fh+gh)+Z(fg+fh+gh)+gh+fg
$$
\vskip-3pt


We will now assume $F=0$. 
We get:

\vskip-5pt
\vskip-5pt
$$
Z(fg+fh+gh)+fg+gh
$$
\vskip-1pt

Due to absence of variables $p$ and $t$ in ${\cal P}$,
this equation FE contains only $Z$ and not $Y$.
FE becomes:
%
$Z(fg+fh+gh)=fg+gh$.
%
Then we need to substitute $Z$ by an expression of type

\vskip-7pt
\vskip-7pt
$$
Z00+Z01*L+Z02*j+Z03*L*j+Z04*h+Z05*L*h+\ldots
$$
\vskip-4pt

\noindent
with the correct 6 variables $L,j,h,f,p,d$ in the correct order.
%
%
FE becomes then:

\vskip-4pt
\vskip-4pt
\begin{verbatim}
 (LZ33+LZ41+Z32+Z40)*dfg +  ... up to ...  (LZ17+LZ21+Z16+Z20)*ghp
\end{verbatim}
\vskip-4pt
From here we obtain a system of simultaneous almost-linear
(except few variables multiplied by $L$) equations starting with:
%


\begin{verbatim}
L*Z33+L*Z41+Z32+Z40=0 ... up to ... L*Z17+L*Z21+Z16+Z20=0
\end{verbatim}

There are many solutions to this equation.
For example when $Z(a,b,c,d,e,f)=a+d+ad+cd+f+af$ our cipher has no linear invariants.
This cipher setting is {\bf not vulnerable to Linear Cryptanalysis (LC)}
in none of the eight cases depending on if $F=0$ or $1$, $L=0$ or $1$ or $K=0$ or $1$.

\vskip-6pt
\vskip-6pt
\subsection{A Major Problem - Simultaneous Solving}
\vskip-2pt

It is possible to see that when $F=1$, the same LZS 317 also has a non-linear invariant
for the same $Z=a+d+ad+cd+f+af$ which is actually of degree 2:
$
{\cal P}=
ef+fg+eh+gh
$ and which is not irreducible.
A ``slight'' problem is that the invariant ${\cal P}$ is {\bf not} the same as when $F=0$. We have not yet broken a block cipher.

We need to simultaneously solve a set of 2 fundamental equations when $F=0$ and $F=1$.
Or a set of up to 8 fundamental equations for different choices of $F,S1,S2$.
We write {\bf up to} 8 because quite frequently some of these are identical which is to our advantage.
Would such a set of equations have a solution at all?

\vskip-6pt
\vskip-6pt
\subsection{Impossibility Results and Provable Security}
\label{SecSecurityProofs}
\vskip-3pt

If this cannot be done,
we would like to be able to prove mathematically
that our $FE$ has no solution and this attack is impossible for our cipher.
There exist some general results of this type in cryptanalysis,
some of which will also apply here
(in particular all those about
anti-invariant properties in
Partitioning Cryptanalysis
cf. \cite{HarpMassThm,
BeiCantResNL,FiliolNotVuln}.
This paper leads indeed to a new well-defined way to {\bf prove} a security of a cipher against
invariant and/or malicious Boolean function attacks,
where one would attempt to prove that FE has no solution.
Such a proof can be done mathematically, or in a more automated way through formal algebra
(a Gr\"{o}bner basis computation showing that there is no solution).
%
A specific way of doing such a proof
which should be easier than in general,
would be to show the impossibility at an early stage,
for example show that elimination of just one variable $F$ is not possible,
cf. Section \ref{ExampleF0F1andL0L01withZ1andZ4Spooky}.
This sort of provable security results could also be used to evaluate the distance
between a secure cipher and one which could be broken in the following way:
for example for AES-128 one could prove that starting from a certain value $N$
no 1-round non-linear invariants which eliminate $N$ key variables exist,
when all the other 128-$N$ key bits 
are fixed.
Then the goal of the cipher designer would be to show that the cipher is secure already for
a very small $N$, which would give a very high level of confidence about the full cipher.



\vskip-9pt
\vskip-9pt
\section{How to Eliminate $F$}
\vskip-6pt

In this section we show a nice example where $F$ gets completely eliminated.
This is an interesting generic or ``milestone'' example as we will see later:
if we can eliminate the constant $F$, 
we can also eliminate a lot more complex things\footnote{
$F$ is a constant known to the attacker.
However {\bf if} $F$ could be totally eliminated
for the purpose of finding a non-linear invariant operating on $X$ bits,
other more complex variables which depend on many key bits and on what happens
in other parts of the cipher, also CAN be 
eliminated, cf.
Section \ref{ExampleF0F1andL0L01preconditionsZ4} below.}.

\subsection{A Construction of a Multiple Invariant}
\label{ExampleF0F1andAnyLexampleQuad1}
\vskip-4pt

We show how an invariant can be constructed in an ordered and systematic way.
We assume that $D(9)=32$ and $D(8)=36$
which implies 
$d  \leftarrow  F + e$ and
 $h  \leftarrow  F + Z1 + a$
and which mandates a sort of imperfect cycle\footnote{
Not a perfect cycle in the sense that it
contains XORs with extra bits
which are not part of the cycle,
cf. later Section \ref{MysteryPtySection}.
}
of length 8 on 8 bits:

\vskip-9pt
\vskip-9pt
\begin{figure}
\vskip-6pt
\hskip-10pt
\hskip-10pt
\begin{center}
\hskip-10pt
\hskip-10pt
\vskip-6pt
\vskip-6pt
\includegraphics*[width=5.1in,height=1.2in]{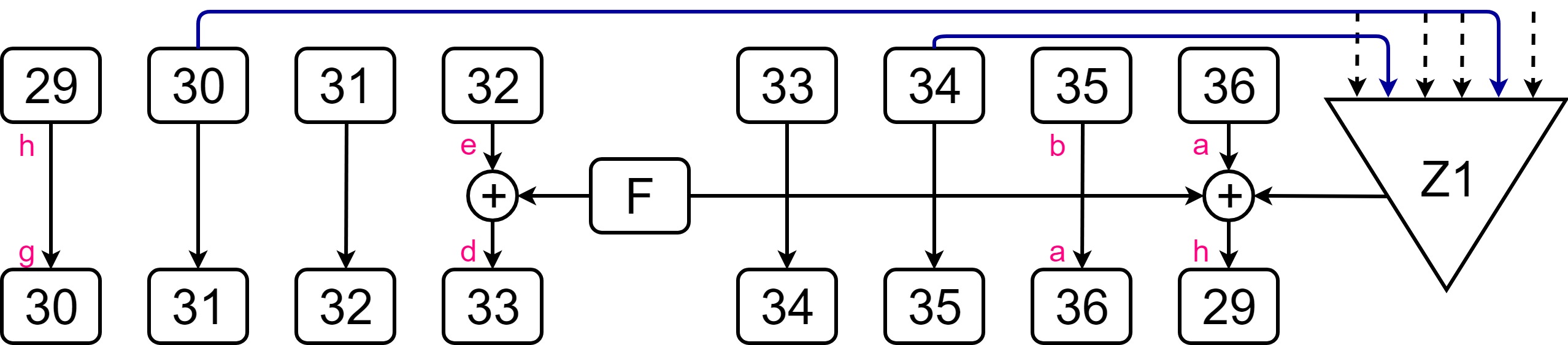}
\end{center}
\vskip-9pt
\vskip-9pt
\caption{
Consequences of assuming that $D(9)=32$ and $D(8)=36$.
$F$ is used twice and will be eventually eliminated.
The intention is that the invariant will not depend on $F$
and neither on 4 additional bits which enter Z1.}
\label{Fig2cycle827Z1proof}
\vskip-9pt
\vskip-9pt
\end{figure}

We will 
decide a bit later (after writing the $FE$)
where different inputs of $Z1$ need to be connected.
We get a series of obvious transitions such as $ce$ becomes $df$
which we would get in traditional Bi-Linear Cryptanalysis (BLC) \cite{BLC},
plus
a series of less obvious transitions due to the
2 assumptions $D(9)=32$ and $D(8)=36$ 
such as $bd$ becomes $ce$, hoping that the term $Fc$
can be somewhat cancelled later.
%

\newpage

\vskip-8pt
\vskip-8pt
\begin{figure}
\vskip-8pt
\hskip-10pt
\hskip-10pt
\begin{center}
\hskip-10pt
\hskip-10pt
\vskip-6pt
\vskip-6pt
\includegraphics*[width=5.1in,height=1.2in]{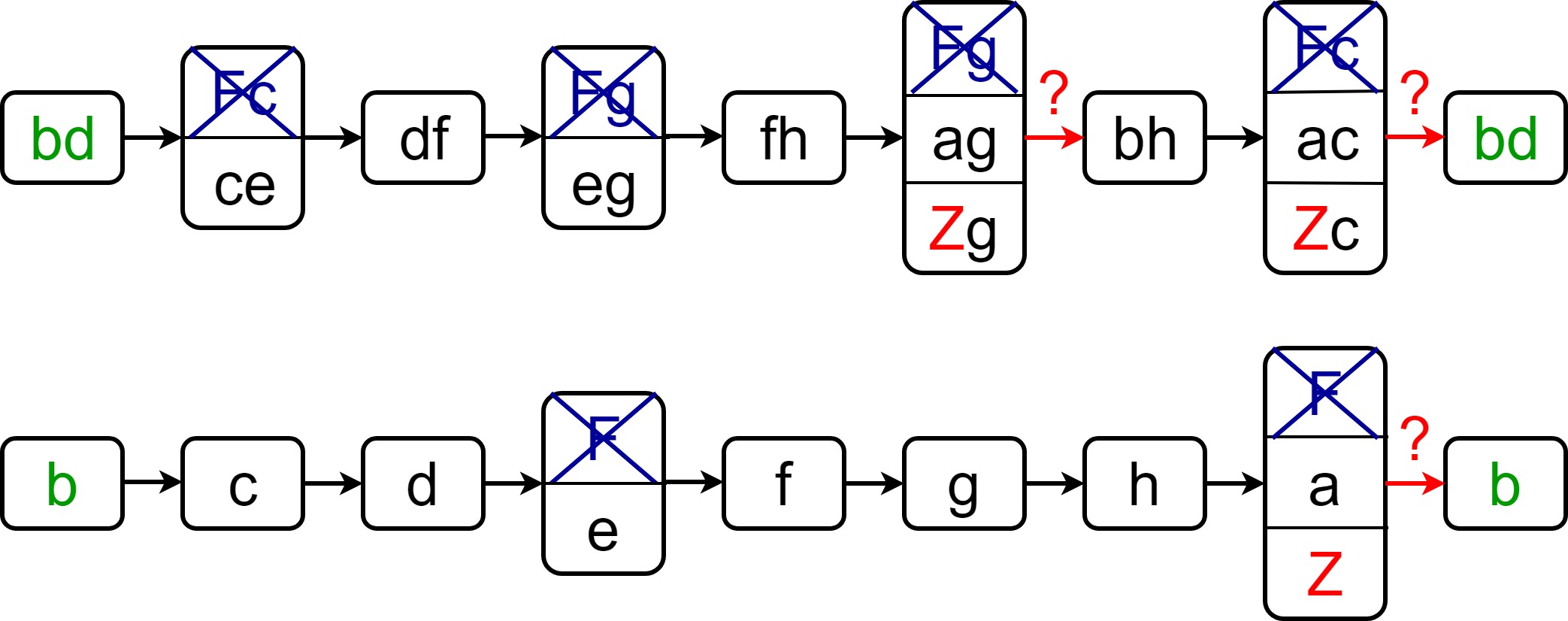}
\end{center}
\vskip-9pt
\vskip-9pt
\caption{A detailed explanation for our invariant which shows terms which cancel.
This analysis is done under our initial ad-hoc assumption that $D(9)=32$ and $D(8)=36$.
}
\label{Fig2cycle827Z1}
\end{figure}
\vskip-9pt
\vskip-9pt

\noindent
This analysis of cycles
on degree 2 monomials suggests to use 
the following irreducible polynomial
defined
as the union of terms in black in both cycles:

\vskip-8pt
\vskip-8pt
$$
{\cal P}=a+b+c+ac+d+bd+e+ce+f+df+g+ag+eg+h+bh+fh
$$
\vskip-4pt

Knowing that each term in blue on Fig. \ref{Fig2cycle827Z1} appears an even number of times,
we have {\bf already eliminated} all terms with $F$.
Now we add all the parts with $Z$ on both cycles,
we expect that our $FE$ will be:
$Z+Zc+Zg\equiv 0$.
This can only work with $Z\not\equiv 0$ if bits $c$ and $g$ are connected as inputs of $Z1$.
This is the moment at which we need to decide which bits will be connected to become inputs of $Z1$,
cf. earlier Fig. \ref{Fig2cycle827Z1proof} and this can be done in any order,
not necessarily following our figure.
For example we can have $P(1)=34$
where $c$ corresponds to $x_{34}$
and $P(4)=30$ where $g$ corresponds to $x_{30}$.
We are now able to generate a long term key
for which our invariant is going to work, for example:

\begin{verbatim}
827: P=34,32,25,30,19,28,18,35,31,33,23,36,24,22,5,1,
13,17,16,10,21,6,20,29,9,15,3 D=21,17,29,24,27,20,31,36,32
\end{verbatim}

For this LZS we can now re-compute our FE which will have fewer unknowns: 

\vskip-5pt
\vskip-5pt
$$
{\cal P}(a,b,c,d,e,f,g,h) =
{\cal P}(b,c,d,F+e,f,g,h,F+Z1+e)
$$
\vskip-3pt


the fundamental equation is then as expected: 

\vskip-4pt
\vskip-4pt
$$
Z=Z(c+g)
$$
\vskip-4pt

and because $FE$ does not depend on either $F$ or $L$,
we do not need eight copies of it but just one.
Here is one solution:

\vskip-4pt
\vskip-4pt
$$
Z=e+be+ce+bce+bf+bcf+bef+bcef
$$



This completes a construction of a non-linear round invariant.
We have checked that there is no linear invariant in any of the eight cases
depending on $F,S1,S2$, and therefore Linear Cryptanalysis (LC) does not work here.
Our {\bf non-linear invariant} ${\cal P}$ works in all eight cases and
therefore it propagates {\bf for an arbitrary number of rounds}
for any key and for any IV.

\newpage 

\vskip-8pt
\vskip-8pt
\subsection{Observations about Our Multiple Invariant}
\label{ExampleF0F1andL0L01observations}
\vskip-4pt

We have obtained an invariant ${\cal P}$ on 8 bits 29-36 the key feature
of which is that it completely eliminates 4 bits which come from
other parts of the cipher: 
This capacity to ignore some bits is crucial in non-linear cryptanalysis because it allows really
to construct relatively simple invariants for very complex ciphers.


\vskip-8pt
\vskip-8pt
\subsection{Pre-Conditions for Our Multiple Invariant}
\label{ExampleF0F1andL0L01preconditions}
\vskip-4pt

In fact the invariant obtained above 
can be constructed systematically.

\setcounter{theorem}{2}
\begin{theorem}[Pre-conditions for Key 827]
\label{Thm827preconditions}
The invariant
${\cal P}=a+b+c+ac+d+bd+e+ce+f+df+g+ag+eg+h+bh+fh$
will occur for $L=0$ or $L=1$,
or for any $L$,
each time the following set of conditions are satisfied:

$$
\begin{cases}
D(9)=32
\cr
D(8)=36
\cr
(1+c+g)Z(L,P[1-5])\equiv 0
\cr
Z\not\equiv 0\cr
\end{cases}
$$
\end{theorem}

\noindent\emph{Proof:}
The constraints $D(9)=32$ and $D(8)=36$ already mandate a cycle between numbers 29-36 shown
on Fig. \ref{Fig2cycle827Z1proof} and they mandate all the transitions of \ref{Fig2cycle827Z1}
which do not have a red question mark (?) sign, which depend on the $FE$.
Finally, we check that all terms in $F$ are eliminated.

\begin{coro}
In particular $Z\not\equiv 0$  implies that two of the $P(1-5)$ values
must be equal to bits 30 and 34, which correspond to letters 'c' and 'g'.
\end{coro}



\vskip-8pt
\vskip-8pt
\subsection{A Transposed Version}
\label{ExampleF0F1andL0L01preconditionsZ4}
\vskip-4pt

Until now we have constructed an invariant for $Z1$ which eliminates $F$
which is known to the attacker (and it also ignores 4 more bits entering $Z1$).
Now 
IF we can eliminate $F$, we can do a lot better.
We can simply {\bf transpose} our invariant from Z1 to Z4 and
here it will eliminate g2, cf. Fig. \ref{FigComplicationUnit6basic} plus another 4 bits
which depend on at least 17 bits and 2 key bits in EXACTLY the same way.
Instead of eliminating a constant $F$ known to the attacker we are now eliminating $g2$ which is a lot more complex to
know, actually it depends on almost everything else
(and the attacker could not possibly know or determine $g2$).

\begin{figure}
\vskip-9pt
\vskip-9pt
\hskip-10pt
\hskip-10pt
\begin{center}
\hskip-10pt
\hskip-10pt
\vskip-6pt
\vskip-6pt
\includegraphics*[width=5.1in,height=1.2in]{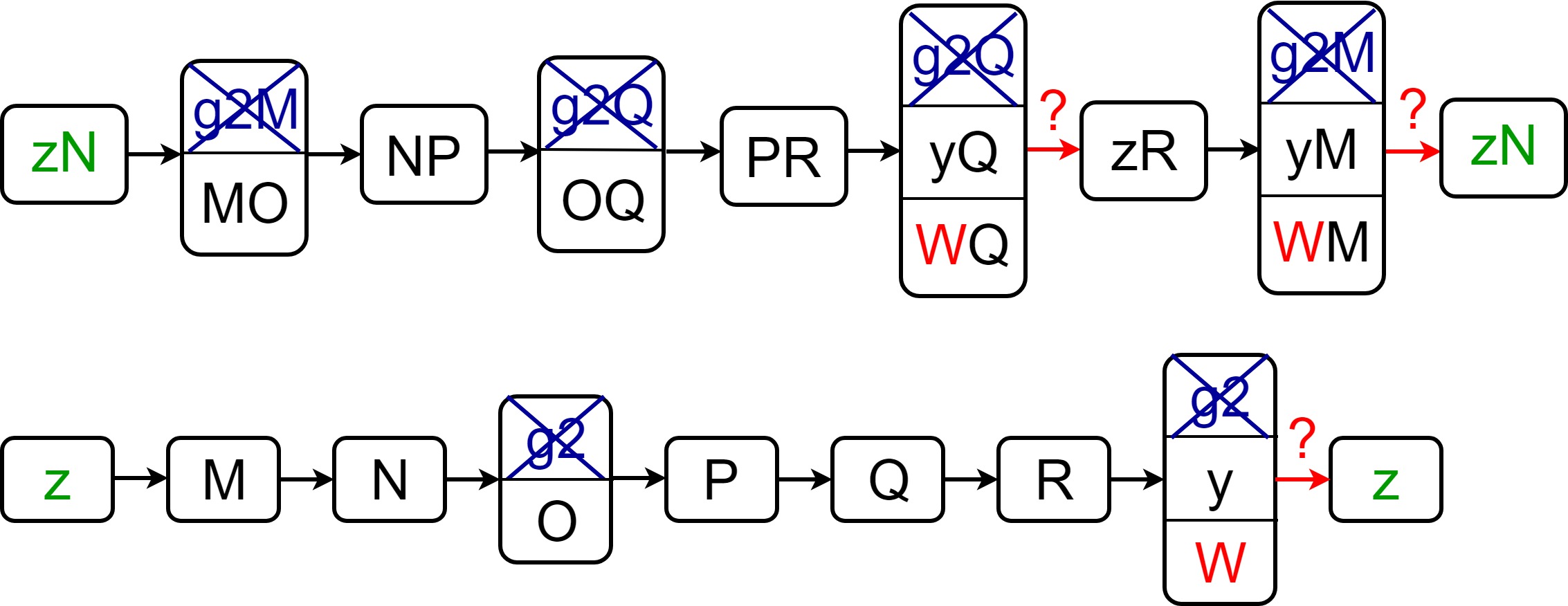}
\end{center}
\vskip-9pt
\vskip-9pt
\caption{The same invariant transposed to $Z4$.}
\label{Fig2cycle847Z4}
\vskip-9pt
\vskip-9pt
\end{figure}

\noindent
By doing so we get a stronger invariant.
For example we found the following LZS:

\begin{verbatim}
847: P=32,22,26,14,21,36,30,17,15,29,27,13,4,23,1,8,35,20,
5,16,24,9,10,6,7,28,12 D=24,12,8,16,36,4,20,28,32
\end{verbatim}

\noindent
Here everything is transposed: the FE is $WM+WQ+W=0$
and is the same in all 8 cases for any $F$ and any $K,L$
and the transposed invariant is now

\vskip-7pt
\vskip-7pt
$$
yM+zN+MO+NP+yQ+OQ+zR+PR+y+z+M+N+O+P+Q+R
$$

This works for a variety of Boolean functions,
and we cannot directly transpose the previous solution
because the two key variables $c,g$ are now $M,Q$ at different positions.
A correctly transposed solution is then for example
$Z=d+cd+bd+bcd+cf+bcf+cdf+bcdf$.
%
A crucial point here is that a 
very complex part of the cipher enters this component at $g2$
and yet it could be totally eliminated.
Furthermore LZS 847 is a permutation on 36 bits 
secure against LC and 
all previously known attacks on T-310 \cite{MasterPaperT310}.



%

\newpage
\vskip-6pt
\vskip-6pt
\section{Construction of Better Simultaneous Invariants}
\vskip-6pt

There are two approaches to this problem: combinatorial and algebraic.
A combinatorial 
approach would be to generate countless examples
and hope that up to 8 equations have a simultaneous solution.
A better -algebraic- approach exists: we will work through intersections of spaces of polynomials
so that we only consider equations where an invariant which works in multiple cases simultaneously can be guaranteed.
Actually eliminating $F$, key bits a.k.a. $K,L$ and what happens in other parts of the cipher is the SAME problem:
we want to construct polynomial invariants which completely ignore a number of bits
which would ruin our cryptanalysis efforts.

\vskip-7pt
\vskip-7pt
\subsection{Is It Possible to Have a Worst-Case Theorem?}
\label{FEReductionToZeroExample2a}
\vskip-3pt

This has been done previously in symmetric cryptanalysis:
In \cite{combmem} we discover that an arbitrarily large number of bits
in an arbitrary complex circuit can be eliminated,
which is guaranteed by a mathematical proof (worst case).
The problem however is,
that the result of \cite{combmem} was obtained under favorable circumstances:
with some redundancy in the number of bits exploitable for the attacker
which can be used to construct the invariant,
including ciphers modified for this purpose.
In this paper we operate in a harder case:
we want to generate invariants on a small number of bits say 14
and yet eliminate say 5 bits per round
which will be key bits or bits coming from other parts of the cipher.
Here
the existence of invariants
will rarely by a worst-case result, rather that the $FE$ happens to have solutions
which sometimes can be guaranteed through specific pre-conditions, e.g. in
Thm. \ref{Thm827preconditions} or Thm. \ref{ThmKT1Cycle9}.

\vskip-9pt
\vskip-9pt
\subsection{Super Strong Invariants or How to Obtain a Worst-Case Result}
\label{FEReductionToZeroExample2b}
\vskip-3pt

Moreover sometimes it actually {\bf can} be guaranteed
that the FE is solvable and that it eliminates $F$ and all the key
bits and all the other bits, and even the main unknown $Z$. 
This happens when ${\cal P}$ is such that $FE$ reduces to 0
(i.e. all its coefficients are 0 for a given ${\cal P}$).
Then it is easy to see that for this invariant,
any Boolean function $Z$ works.
This does not say if some polynomials ${\cal P}$ where FE actually reduces to 0 would exist.
They do exist, cf. Appendix \ref{SimpleInvP20Cycle9BiasedFEreducedto0App}. 

This sort of invariant is considered out of the scope for the present paper
as typically they impose too many constraints on LZS and attack would
work for very few LZS.
Moreover they are typically quite trivial
and are not secure against LC 
cf. Appendix \ref{SimpleInvP20Cycle9BiasedFEreducedto0App}.
Yet 
they work for Boolean functions which satisfy very strict criteria
and also for the original Boolean function $Z$
used in 1980s. 
Here 
no Boolean function can make this cipher
setup secure against round invariant attacks
(and against partitioning attacks in general).



\vskip-8pt
\vskip-8pt
\section{Simultaneous Invariants through Intersection of Spaces}
\label{MultipleInterestectElimF}
\vskip-5pt

In this section we show that the question of simultaneous solvability of FE for several cases
can be studied in terms of row-echelon elimination
of specific variables in a set of multivariate polynomial equations.
Previously we have eliminated $F$ plus 4 more bits and then transposed this result
to eliminate a quantity 
with even more complex dependencies.
Here we go one step further.

\vskip-9pt
\vskip-9pt
\subsection{Construction of Invariants on Two Remote Parts of the Cipher}
\label{ExampleF0F1andL0L01withZ1andZ4Spooky}
\vskip-4pt

We are going to show the existence of an invariant
on 
$Z1$ and $Z4$ which 
mixes bits which sit at two opposite ends\footnote{
We call it ``spooky interaction''
the two distant (as remote as only possible) parts of the cipher ``talk'' to each other
in terms of a polynomial invariant which mixes variables from both sides.
However their principal connection a.k.a. 
$g27$ is eliminated.}
 of the cipher, cf. Fig \ref{FigComplicationUnit6basic}.
These parts are connected through $Z2$ and $Z3$ by a quantity called $g27$ and defined as
$g27=g2+g7$ cf. Fig. \ref{FigComplicationUnit6basic}
which depends 
on an excessively large number of round input bits (at least 19)
plus the key bit $S2$.   
Without $g27$ none of the outputs on the right hand side we use can be computed.
Yet this connection $g27$ 
gets totally eliminated (so does $F$ and few other things).
This leads to an invariant property which does NOT depend on neither key bits
nor on
what happens in other
parts of the cipher.

\vskip-9pt
\vskip-9pt
\subsection{Construction of an Invariant with Z1 and Z4}
\label{ExampleF0F1andL0L01withZ1andZ4Spooky}
\vskip-4pt

We are going now to show how such invariants can be constructed essentially from scratch,
however aiming at connecting $Z1$ and $Z4$ specifically.

\vskip-9pt
\vskip-9pt
\begin{figure}
\vskip-6pt
\hskip-10pt
\hskip-10pt
\begin{center}
\hskip-10pt
\hskip-10pt
\vskip-6pt
\vskip-6pt
\includegraphics*[width=4.9in,height=2.6in]{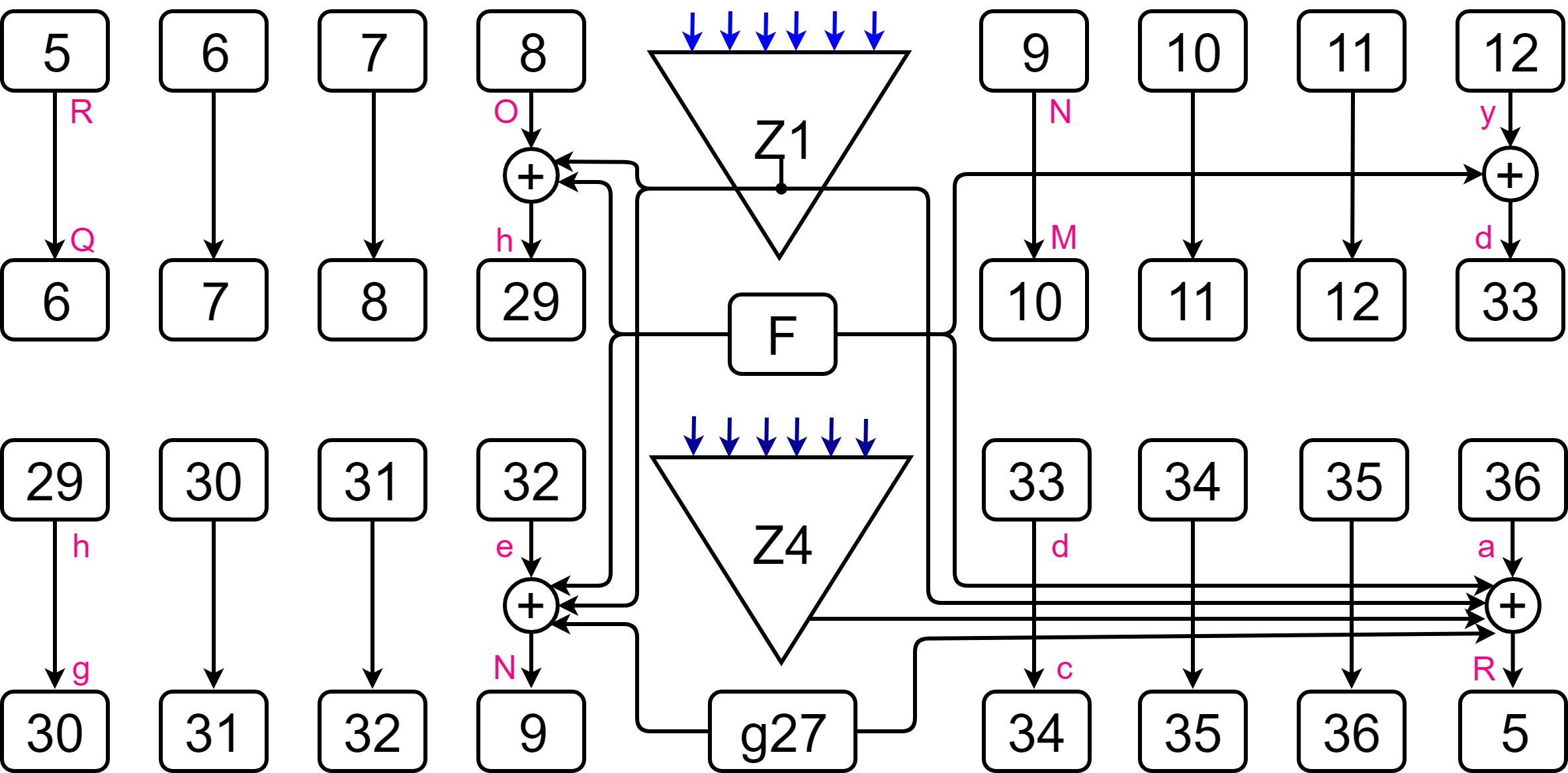}
\end{center}
\vskip-9pt
\vskip-9pt
\caption{
Our ad-hoc assumption is
$D(2)=36$, $D(3)=32$, $D(8)=8$, $D(9)=12$.
Here $g27\stackrel{def}{=}g2+g7$ cf. Fig. \ref{FigComplicationUnit6basic}.
We aim at invariants on 8+8 bits which would not depend on
$F,g27$ and few more inputs of $Z1,Z4$ in blue are not yet connected.
Other inputs of $Z1,Z4$ will be used and
their connections are decided at a later stage.
}
\label{Fig4cycle714Z1Z4}
\end{figure}
\vskip-4pt
\vskip-4pt
\vskip-4pt


We focus on 8+8 bits 29-36 and 5-12 pertaining to $Z1$ and $Z4$ only
and strictly avoiding anything between g2 and g7.
We start by assuming
the following four constraints
 which implements a basic sort of ``exchange'' connection between two opposite ends of the cipher.
It also implements a cycle of size 4 on the $D(i)$ and a cycle of size 16 on the
input/output variables, which we intend to be used in our attack,
and the whole could be viewed as 4 shift registers connected in a loop,
cf. Fig.
\ref{FigComplicationUnitSKSTotalOrderT-310Case} page
\pageref{FigComplicationUnitSKSTotalOrderT-310Case}.

\vskip-9pt
\vskip-9pt
$$
\begin{cases}
D(2)=4\cdot 9\cr
D(3)=4\cdot 8\cr
D(8)=4\cdot 2\cr
D(9)=4\cdot 3\cr		
\end{cases}
$$
\vskip-4pt



We can again generate cycles
in the same way as in Section \ref{ExampleF0F1andAnyLexampleQuad1}:
transitions are either natural  e.g. $bc\leftarrow cd$
or consequences of the 4 conditions on $D()$ above.
Similarly we ignore the boxes with blue crosses which we hope might eventually 
be eliminated later inside the final $FE$ which is not yet finalized.
A detailed analysis of these natural cycles
as shown on Fig. \ref{Fig4cycle714Z1Z4} (and few more)
leads to 8 natural
clusters of monomials for ${\cal P}$ which most likely work together:

\vskip-5pt
\begin{figure}
\vskip-8pt
\vskip-8pt
\vskip-8pt
\vskip-8pt
\hskip-6pt
\hskip-6pt
\begin{center}
\hskip-6pt
\hskip-6pt
\includegraphics*[width=5.1in,height=3.2in]{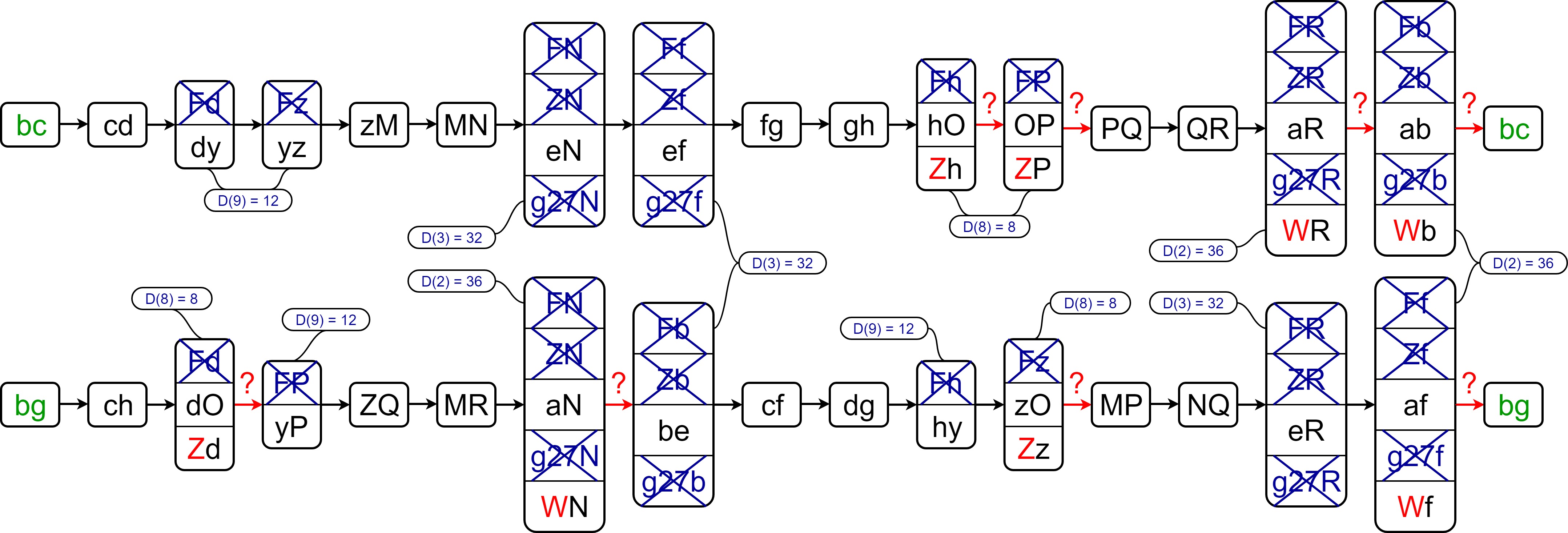}
\end{center}
\vskip-9pt
\vskip-9pt
\caption{A detailed analysis of transitions we aim at using in our invariant.
The boxes with crosses are terms we hope to cancel later.
Transitions in red with ? depend on $Z$ and will eventually work
only if our final FE equation has a solution.}
\label{Fig4cycle714Z1Z4cycles}
\vskip-6pt
\vskip-6pt
\end{figure}
\vskip-1pt


\vskip-8pt
\vskip-8pt
\subsection{Analysis of Polynomial Spaces}
\label{ExampleF0F1andL0L01withZ1andZ4Spaces}
\vskip-4pt

This gives 8 natural pre-FE equations, 
with the idea that a final ${\cal P}$ and final $FE$ is a fixed linear combination of these $8$,
which are exactly:

\vskip-5pt
\vskip-5pt
{\scriptsize
\begin{verbatim}
Wb+Fd+(b+f+N+R)(F+Z+L+Y+X+P6+P13+P20)+h(F+Z)+Fz+ P(F+Z)+WR
(L+Y+X+P6+P13+P20)(c+g+M+Q)+c(Z+W)+ZM+WQ
Fb+FN+W(d+P)+ (F+Z)(f+R)+ (F+Z+L+Y+X+P6+P13+P20)(d+h+z+P)
F(Z+W+a+e+y+O)+We+Zy+(Z+L+Y+X+P6+P13+P20)(W+a+e+1)
Z(d+z)+W(f+N)+F(d+h+z+P)+(F+Z+L+Y+X+P6+P13+P20)(b+f+N+R)
W(g+M)+Z(g+Q)+(L+Y+X+P6+P13+P20)(c+g+M+Q)
Z(b+N)+W(h+z)+F(b+f+N+R)+(F+Z+L+Y+X+P6+P13+P20)(d+h+z+P)
ZW+Z(a+e)+W(y+O)
\end{verbatim}
}
\vskip-6pt

\noindent
We start with a polynomial space of dimension 8.
Now we observe that we need to eliminate $Fb$ and other similar monomials.
A standard row-echelon procedure forces us to XOR some equations
and we obtain the following $6$ linearly independent equations:

\vskip-5pt
\vskip-5pt
{\scriptsize
$$
\begin{cases}
XR+WR+(R+h+z+P+N+f+b+d)(P6+P13+P20)+Zb+Lb+\cr ~~~~
Xb+Lf+Yf+Xf+Lh+Yh+Xh+Zz+Lz+Yz+Xz+LP+YP+XP+\cr ~~~~
WP+Yb+YN+XN+LR+YR+Zd+Ld+Yd+Xd+Wd+ZN+Wb+LN\cr
Zc+Lc+Yc+Xc+Wc+(c+g+M+Q)(P6+P13+P20)+\cr ~~~~
Lg+Yg+Xg+ZM+LM+YM+XM+LQ+YQ+XQ+WQ\cr
WR+Wf+WN+Zh+Zz+ZP+Zd+Wb\cr
Lc+Yc+Xc+(c+g+M+Q)(P6+P13+P20)+Lg+Yg+Xg+\cr ~~~~
+LM+YM+XM+LQ+YQ+XQ+Zg+Wg+WM+ZQ\cr
Zb+Zf+WP+ZR+Wd+Wh+Wz+ZN\cr
ZW+Za+Ze+Wy+WO
\end{cases}
$$
}
\vskip-3pt

This stage is crucial for cipher designers, possibly one should be able to construct a block cipher in such a way that the dimension of this set of
polynomials is already $0$ [here it is $6$ so we can do cryptanalysis].
Then we are going to eliminate all products of $P6$ in the same way, which also leads to elimination of numerous other monomials
and this leads to a dimension $3$, still not 0:

\vskip-3pt
\vskip-3pt
$$
\begin{cases}
Zc+Wc+ZM+WQ+Zg+Wg+WM+ZQ\cr
WR+Wf+WN+Zh+Zz+ZP+Zd+Wb\cr
ZW+Za+Ze+Wy+WO\cr
\end{cases}
$$
\vskip-3pt

Finally it is possible to show that out of $2^{3}-1=7$ possible linear combinations of these ${\cal P}$,
only the first $2$ out of $7$ lead to solutions,
this under the condition that the cipher wiring $P()$ is injective,
or we get FE equations for which the set of solutions is empty due to the simple fact
that a Boolean function cannot annihilate variables which are not inputs of this function.
This leads to only 2 possibilities,
out of which we have chosen to work with one:

\vskip-8pt
\vskip-8pt
$$
{\cal P}=bc+cd+dy+yz+zM+MN+eN+ef+fg+gh+hO+OP+PQ+QR+aR+ab+bg+ch+
$$
\vskip-9pt
\vskip-9pt
\vskip-9pt
$$
dO+yP+zQ+MR+aN+be+cf+dg+hy+zO+MP+NQ+eR+af
.
$$
\vskip-3pt

\noindent
This ${\cal P}$ is irreducible and the FE is obtained by repeating the very same linear transformations
by which we have reduced the dimension from 8 to 2 above:

\vskip-3pt
\vskip-3pt
$$
NW+PZ+RW+Wb+Wf+Zd+Zh+Zz
$$
\vskip-3pt


All we have now to do is to ensure through $P()$ that various inputs
which appear in the $FE$ above are connected to
$Z1=Z$ or $Z4=W$ respectively,
for example $f$ must by an input of $W$,
therefore we need $P(i)=31$ for some $i\in\{21,\ldots,26\}$.
This is exactly the moment at which we can decide all the connections in blue on
Fig. \ref{Fig4cycle714Z1Z4}.
Here is an example of a LZS 
where this ${\cal P}$ 
works: 

\vskip-3pt
\vskip-3pt
\begin{verbatim}
714: P=11,7,30,29,33,1,20,17,2,15,14,27,36,24,18,8,19,
23,28,32,4,16,31,9,35,5,13 D=16,36,32,24,4,28,20,8,12
\end{verbatim}
\vskip-3pt


\noindent
This example was found by the exact steps we enumerate below and by feeding
the resulting set of constraints on $D()$ and $P()$ to a SAT solver at the end.
This is done many times until we find a valid permutation on 36 bits.

\noindent
{\bf Solving the FE.}
It remains to find a solution to $NW+PZ+RW+Wb+Wf+Zd+Zh+Zz\equiv 0$.
There is still some degree of freedom here in selecting which bits will be inputs
of $W$ and $Z$ though function $P()$.
With LZS 714 above, one possible solution is
\vskip-4pt
\vskip-4pt
$$
Z=1+dc+cb+fb+b+c+de+df+db+e+f+d+eb
.$$
\vskip-4pt
\noindent
We get another invariant which works
for any number of rounds and any key.


\vskip-8pt
\vskip-8pt
\subsection{An Invariant with Z1, Z2 and Z4}
\label{ExampleF0F1andL0L01withZ1Z2Z4}
\vskip-4pt

In the same way
we have constructed another yet more complex invariant
which involves three Boolean functions Z1,Z2 and Z4
(and yet it deliberately avoids $Z3$ and all the 
120 secret key bits of type $S2$).
We start with the following
cycle of size 6 
(the size of the cycle increases and we are getting closer the maximum possible size,
cf. Fig.
\ref{FigComplicationUnitSKSTotalOrderT-310Case} page
\pageref{FigComplicationUnitSKSTotalOrderT-310Case}):

\vskip-3pt
\vskip-3pt
$$
\begin{cases}
D(2)=4\cdot 9\cr
D(3)=4\cdot 8\cr
D(6)=4\cdot 3\cr
D(7)=4\cdot 2\cr
D(8)=4\cdot 7\cr
D(9)=4\cdot 6\cr		
\end{cases}
$$
\vskip-4pt

We repeat the same steps as before
which leads to a polynomial space of dimension 12 (instead of 8).
After elimination of all monomials in $F$ and $P6$ etc.
the dimension is still not $0$, and we get the following plausible polynomial:

\vskip-8pt
\vskip-8pt
$$
{\cal P}=bl+cO+dP+mQ+nR+ao+bp+cy+dz+mM+nN+eo+
$$
\vskip-5pt
\vskip-5pt
\vskip-5pt
$$
fp+gy+hz+iM+jN+ek+fl+gO+hP+iQ+jR+ak
$$

and the $FE$ is

\vskip-7pt
\vskip-7pt
$$
Yc+Yg+Wk+Wo+ZM+ZQ
$$
\vskip-3pt

This $FE$ allows to see that each of the three Boolean functions needs two specific
variables listed here as inputs, which we need to mandate using $P()$.
With all the constraints above we can now construct a possible LZS key:

\vskip-7pt
\vskip-7pt
\begin{verbatim}
124: P=6,36,5,18,10,16,24,12,20,34,30,31,7,21,13,11,
23,35,1,32,26,2,4,22,8,28,9 D=0,36,32,20,4,12,8,28,24
\end{verbatim}
\vskip-4pt

\noindent
Finally, after fixing the exact placements of different inputs of $Z$, $W$ and $Y$,
and not earlier, we can find a common solution for the $FE$. 
This is less trivial than before because we have 
three instances of the SAME Boolean function, see
Def. \ref{defiRewriteFEZ00} page \pageref{defiRewriteFEZ00},
and the solution depends on how the 3 sets of 2 bits are mapped onto 6 bits.
We selected an example LZS124 which is not quite trivial, and the solution is:
$
Z=1+dfe+ba+fa+dbe+fae+db+f+b+df+ad+ae+d+de+fad+fe+e+a+eb+bad+bae
$.

It is possible to see that unlike in previous examples no solution $Z$ of degree 2 exists.
We see that as our examples become more complex, the complexity of $Z$ also increases.
This is needed if eventually we wanted to construct a weak cipher where $Z$ would be
a Boolean function with no apparent weakness.


\vskip-9pt
\vskip-9pt
\section{From Invariants to Ciphertext-Only Attacks on T-310}
\vskip-6pt
\label{CiphertextOnlyAttacks}

Once we are able to construct a certain polynomial invariant property for our block cipher
breaking the cipher for at least some [weaker] LZS keys is not too difficult\footnote{
There are however some serious pitfalls in this process,
cf. Section \ref{CiphertextOnlyAttacksTrivialCases} and App. \ref{SimpleInvP20Cycle9BiasedFEreducedto0App}.
}.
The key observation cf.
\cite{JakobsenPartitioning} is that typically our polynomial invariants ${\cal P}$
will lead to partitioning the space of say $2^{16}$ elements into two sets
of rather {\bf unequal} sizes.
Our experience shows that the sizes are frequently not equal with some strong bias(!).
More frequently than not\footnote{
This works for the invariants of Section \ref{CiphertextOnlyAttacksTrivialCases}
and  Section \ref{ExampleF0F1andL0L01withZ1andZ4Spaces}.
Now if partitions are of equal sizes,
cf. Section \ref{SimpleInvP20Cycle9Balanced}, the correlation attack we outline here will not work.}
as we are going to show below,
we are able to produce {\bf a strong pervasive bias}
on the state of our block cipher which carries for any number of rounds.

\subsection{How to Exploit a Non-Linear Invariant: Higher Order Correlations}
\vskip-6pt
\label{CiphertextOnlyAttacksGoodCases}

We first  consider the invariant from Section \ref{ExampleF0F1andL0L01withZ1Z2Z4}:

\vskip-8pt
\vskip-8pt
$$
{\cal P}=bl+cO+dP+mQ+nR+ao+bp+cy+dz+mM+nN+eo+
$$
\vskip-7pt
\vskip-7pt
\vskip-7pt
$$
fp+gy+hz+iM+jN+ek+fl+gO+hP+iQ+jR+ak
$$
\vskip-4pt

which is irreducible\footnote{Otherwise
we could be dealing with a multiple linear attack, see
Appendix \ref{SimpleInvP20Cycle9BiasedFEreducedto0App}.}, has 24 variables
and a computer simulation shows that the ${\cal P}=0$ in
exactly
$
8519680
$
cases instead of 8388608 expected.

Even though any individual variable say $a$ or $N$ is typically not biased,
neither are pairs of variables,
we observed that in each case there exists a relatively small $N$
such that for ANY subset of $N$ out of 24,
the joint probability distribution of these $N$ variables in not uniform.
Moreover this $N$ is not excessively large in practice, for example the reader can easily verify
that the event $abcdef=1$ AND ${\cal P}=0$ for the above polynomial happens 0 times out of 133120 expected,
which demonstrates an extremely strong bias. Yet $N=6$ is yet rather high.

\vskip-6pt
\vskip-6pt
\subsection{More Concrete Examples of Higher Order Correlations}
\vskip-6pt
\label{CiphertextOnlyAttacksGoodCases}

Another good example is the invariant of Section \ref{ExampleF0F1andL0L01withZ1andZ4Spaces}
where we have the partition in two sets with $36864$ and $28672$ elements,
and we have also checked that it produces a serious bias with $N=5$.
For example when ${\cal P}=0$ yet the event $abcde=1$ happens 1280 times
and the event $abcd(e+1)=1$ happens 1024 times.
Moreover the event $abcdef=1$ never happens.
In addition we have also checked that the polynomial ${\cal P}$ is irreducible
and that there are no linear invariants true with probability 1 and therefore
we really found a non-trivial attack.

In Section
\ref{SimpleInvP20Cycle9BiasedFEHomQuad} we present a better example
working already with $N=3$.

\vskip-6pt
\vskip-6pt
\subsection{A Higher Order Correlation Attack}
\vskip-6pt
\label{CiphertextOnlyAttacksHOCorrelation}

From here we can construct a strong correlation attack
where the bias does {\bf not} depend on the number of rounds as follows.
We assume that $N=5$ produces a bias as above.
The value $N=5$ if probably sufficient in practice.
Now it is sufficient that some Boolean function say Z3 in the next round\footnote{
$Z3$ is presently NOT constrained at all
by the necessity for ${\cal P}$ to be an invariant.}
takes some subset of 5 bits out of our 24 as an input.
It can be any subset, so that likelihood this will actually happen is high.
Then Z3 will take 1 more bits (presumably not one of the 24) and with a high probability,
the output of this Boolean function will be biased 
at {\bf every} encryption round.
This will work if sufficiently many of the 24 bits are used with $Z3$ or with $Z2$
(which in turn depends on how many bits in $Z1$ and $Z4$ come from $Z3$ and $Z2$, 
which influence we were trying hard to maximize!).
Such permanently biased bits will inevitably lead to correlation attacks where
$S1$ and the special output bit $x_{\alpha}$ actually used in the encryption \cite{MasterPaperT310}
will be connected through just a few biased bits.
This, given the fact that the same key
bit $S1$ is repeated every 120 rounds, will inevitably lead to a ciphertext-only correlation attack on T-310 and
on a key recovery attack on 120 bits of $S1$ key (recovering $S2$ could be harder).
Knowing that bits in any natural language (not only German) in any reasonable encoding
are {\bf always} strongly biased, cf. \cite{CiphOnlyKT12ucry18}, from any such a bias we can
infer concrete values of individual key bits $S1$ by majority voting.
%
We obtain an attack of the sort which is extremely rare in cryptanalysis:
a correlation attack with recovery for at least 120 bits of the key,
where the correlation does {\bf NOT} degrade as the number of encryption rounds increases.
It is also quite unique in another way:
the biases which come from the non-linear invariant
are higher order biases on joint probability distributions of a certain dimension $N$.
They appear almost ex-nihilo,
and they are NOT biases which could be produced or understood within the
strict framework of Linear Cryptanalysis.


\subsection{Important Remark - Avoiding Trivial Cases}
\vskip-6pt
\label{CiphertextOnlyAttacksTrivialCases}

It is important to see that NOT every
partitioning in two spaces of unequal sizes will work here.
For example what we propose above will
NOT work many of the (so called stronger) invariants according to Section \ref{FEReductionToZeroExample2b}.
An eminent example of this is given in Appendix \ref{SimpleInvP20Cycle9BiasedFEreducedto0App} where ${\cal P}=0$ in 3/4 
entirely due to Linear Cryptanalysis.
The ultimate test is to see 
if a some set of $N$ variables exhibits a bias.
There are plenty of trivial polynomials in non-linear cryptanalysis which
are products of linear factors and come from standard Linear Cryptanalysis (LC) with multiple linear approximations,
cf. Appendix \ref{SimpleInvP20Cycle9BiasedFEreducedto0App} and Section 21 in \cite{MasterPaperT310}.
Making sure that we are not in this case is however easy.
Most of the time it will be sufficient to check if the polynomial is irreducible
and if not, we expect to find some biases (there will be however exceptions to this rule).
It is also good when it is not symmetric,
and it is useful if it contains no variables in linear parts which would not appear in non-linear parts
(otherwise it is likely to be balanced).


\section{The Mystery Property and Invariants with $P(20)$}
\label{MysteryPtySection}
\label{ThmKT1Cycle9InsideSection9}
\vskip-5pt

In this section we
attempt to elucidate an old Cold War mystery question. 
There exists an extensive 123 page ``master'' document on T-310 from 1980,
where we read that the round function of this cipher must be bijective \cite{T-310An80}.
This property however is {\bf not} required for encryption given that the cipher is used in a stream cipher mode.
Detailed mathematical proofs of this fact can be found in \cite{CiphOnlyKT12ucry18}
which paper also shows how to break this cipher in the otherwise case,
which would explain why the designers mandated this property.
We have studied the proofs in \cite{CiphOnlyKT12ucry18} and compared it to the whole
of (extremely complex) specification of the KT1 class of LZS \cite{T-310An80},
concerning 
about 90 $\%$ of keys used in real-life state communications \cite{MasterPaperT310}.
The following result appears as Thm. 5.1. in \cite{LCKT1ucry18}:

\vskip-6pt
\vskip-6pt
\setcounter{theorem}{0}
\begin{theorem}[KT1 Cycling Theorem]
\label{EveryKT1CycleThm}
For every key in the class KT1
if we replace the first value $D(1)$ by
$P(20)$ and we divide all values by 4,
we obtain a permutation
of the set $\{1,\ldots,9\}$ with exactly one cycle.
\end{theorem}
\vskip-4pt

\begin{figure}
\hskip-6pt
\hskip-6pt
\begin{center}
\hskip-6pt
\hskip-6pt
\includegraphics*[width=5.1in,height=1.3in]{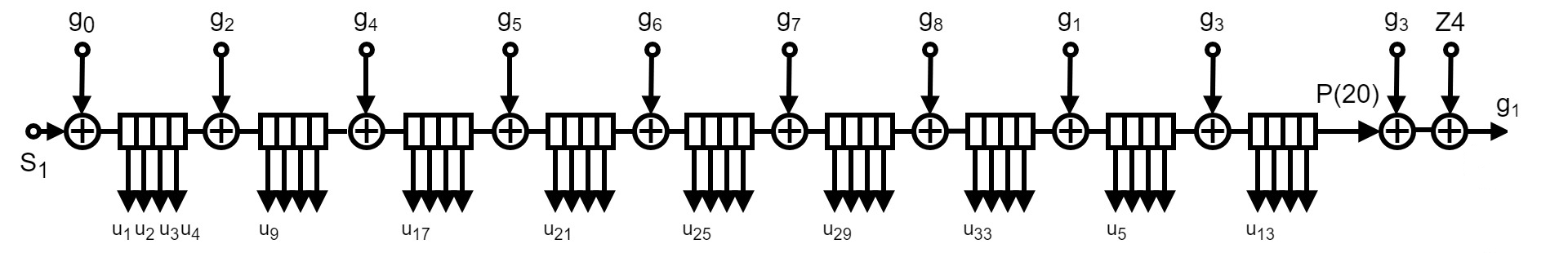}
\end{center}
\vskip-3pt
\vskip-6pt
\caption{
A typical ``serial connection of shift registers'' situation
inside T-310 for a long-term key of type KT1
cf. Thm. \ref{EveryKT1CycleThm}.
The starting point is the key bit S1 (due to $D(1)=0$).
We also show what happens 
after the last bit re-enters the cipher at $P(20)$.
}
\label{FigComplicationUnitSKSTotalOrderT-310Case}
\vskip-9pt
\vskip-9pt
\end{figure}

\noindent
Yet this property is NOT used nor it is required in the KT1 security proof of \cite{CiphOnlyKT12ucry18}.
If so, what is this property for? Does it serve a purpose?
We call it a ``mystery'' property.
In fact, it has a lot to do with our attacks.
The main point is that given the specific Feistel cipher structure in which each freshly created bit is used exactly 4 times,
absolutely every of our invariants here and also all 20+ pages of linear invariants constructed in \cite{MasterPaperT310}
can be seen as acting on some set of  $Di$ and bit $P(20)$ (or e.g. $P(27)$) in a specific order
eventually forming a closed cycle.
All invariants here and elsewhere \cite{MasterPaperT310,LCKT1ucry18} can be analysed in this way.
%
%
%
Such cycles are inevitable, and exist in {\bf every} block cipher
and they do NOT mean that a specific cipher setup is weak.
This is because 
the bits are on the way also XORed with many other bits,
cf. bits denoted as $g_i$ on Fig. \ref{FigComplicationUnitSKSTotalOrderT-310Case}
and on Fig. \ref{FigComplicationUnit6basic},
which depend on countless other bits in a very complex way.
Only sometimes we have some invariant attacks which eliminate these $g_i$ or some other bits\footnote{
The whole point about all our invariants it to use a restricted set of bits and
eliminate SOME, not all of the $g_i$, which can include elimination of whole intervals
inside 
Fig. \ref{FigComplicationUnitSKSTotalOrderT-310Case},
e.g. in Section \ref{ExampleF0F1andL0L01withZ1andZ4Spooky},
and eliminating also (preferably all!) the key bits $S1$ and $S2$,
in the same way as our previous attacks avoided $F,g2,g27,Z3,S2$ etc.}.

\vskip-8pt
\vskip-8pt
\vskip-8pt
\vskip-8pt
\vskip-4pt

\vfill
\newpage

\vskip-8pt
\vskip-8pt
\subsection{Short Cycles and Linear Cryptanalysis}
\vskip-4pt

This question of cycles already arises 
in Linear Cryptanalysis.
We recall that our cipher shift bits of the form $1\to 2\to 3\to 4$ and then bit $4$ is no longer used [it is erased]
or $25\to 26\to 27\to 28$.
Some cases are quite weak, for example if we had $d[9]=0$
we have $33\to 34\to 35\to 36$ and $36+S1\to 33$,
and we get a simple linear invariant with one key bit $S1$.
A more complex example is Thm. J.1. in \cite{MasterPaperT310} which states that
for some proportion of all LZS
we have the following linear approximation 
[9,13]$\to$[10,14]$\to$[11,15]$\to$[12,16]$\to$

\noindent
[25,29,33]$\to$[26,30,34]
$\to$[27,31,35]$\to$[28,32,36]$\to$[9,13].
Here Thm. J.1. in \cite{MasterPaperT310} states that
this linear approximation will occur when we have for example \\
$(P(20)/4,D(3)/4,D(4)/4,D(7)/4,D(9)/4=(7,9,8,3,4)$.
We have a permutation on the set $\{3,4,7,8,9\}$ if $P(20)$ is identified with 8.
Interestingly this cycle and  Thm. J.1. of \cite{MasterPaperT310}
do {\bf not} contradict our Thm. \ref{EveryKT1CycleThm} above.
This is because instead of replacing $D(0)$ by $P(20)$ we replace $D(8)$ by $P(20)$ in Thm. J.1. cf. \cite{MasterPaperT310}.
Moreover Thm. J.1. of \cite{MasterPaperT310} shows more than one way in which this can be made to work
in spite of Thm. \ref{EveryKT1CycleThm} and few more similar theorems can be found in \cite{MasterPaperT310}.
Overall we see that existence of invariant attacks does not  contradict Thm. \ref{EveryKT1CycleThm}.
In Thm. J.1. of \cite{MasterPaperT310} we have a cycle of length 5
and the invariant is trivial (completely linear).
Can we find a non-linear invariant with a cycle of length 9 and with the exact conditions of Thm. \ref{EveryKT1CycleThm}?
Yes we can and two 
examples will be shown below. 
In general given that Thm. \ref{EveryKT1CycleThm} does not prevent linear attacks cf. \cite{MasterPaperT310},
it should not prevent their non-linear generalizations either(!).
On the contrary, we believe that Thm. \ref{EveryKT1CycleThm} helps:
to construct invariant properties of a very specific shape.
Moreover it is not true either,
that (at
least) Thm. \ref{EveryKT1CycleThm} would prevent invariants which would avoid
some key bits, cf. for example Thm. J.1. and also LZS 778 in Table 23 p. 87 in \cite{MasterPaperT310}.
As a proof of concept
we
are now going to
construct specific invariants with KT1 keys
which therefore will be totally compliant with Thm. \ref{EveryKT1CycleThm}. 

\vskip-6pt
\vskip-6pt
\subsection{An Invariant Using $P(20)$ and Working for a KT1 Key}
\label{SimpleInvP20Cycle9KT1}
\label{SimpleInvP20TwoProducts}
\vskip-2pt

We present a 
proof of concept with a full cycle of length 9 and $P(20)$.
We have first with paper and pencil imagined $D$
with a cycle of length 9 compliant with Thm. \ref{EveryKT1CycleThm}
and the simple invariant with 11 variables: 

\vskip-5pt
\vskip-5pt
$$
{\cal P}=an+gn+u+v+w+x+O+P+Q+R
$$
\vskip-1pt

Then we assume 
$D(2)=36$, $D(4)=8$ and few more constraints cf. Thm. \ref{ThmKT1Cycle9} below,
aiming at the following sort of configuration: 


\vskip-3pt
\vskip-3pt
\setcounter{theorem}{2}
\label{ThmKT1Cycle9}
\begin{theorem}[KT1 quadratic invariant construction framework]
For each long term KT1 key s.t. $D(2)=36$, $D(4)=8$, $P(20)=16$, 
$\{22,23,30,35,36\} \subset \{P(21),P(22),P(23),P(25),P(26)\}$,
if the Boolean function 
is such that $W = a+an+bo+gn+ho$,
and for any short term key on 240 bits, and for any initial state on 36 bits,
we have the non linear invariant, $\mathcal{P}=an+gn+u+v+w+x+O+P+Q+R$, which is true with probability exactly 1.0 for 1 round.
\end{theorem}
\vskip-2pt

\noindent\emph{Proof:}
We need to distinguish variables such as $P$ at the input called $P_0$
and $P$ at the output called $P_1$. We can successively XOR the expression
$a_0n_0+g_0n_0+u_0+v_0+w_0+x_0+O_0+P_0+Q_0+R_0$ with $W+a_0+a_0n_0+b_0o_0+g_0n_0+h_0o_0$
and $R_1+g3+u_0+W+a_0$ and $x_1+g3+O_0$ and we do all the trivial steps like $v_0\leftarrow u_1$
and get the final result $a_1n_1+g_1n_1+u_1+v_1+w_1+x_1+O_1+P_1+Q_1+R_1$.

\vskip-8pt
\vskip-8pt
\begin{figure}
\vskip-8pt
\vskip-8pt
\vskip-8pt
\hskip-8pt
\hskip-8pt
\begin{center}
\hskip-8pt
\hskip-8pt
\includegraphics*[width=5.1in,height=2.4in]{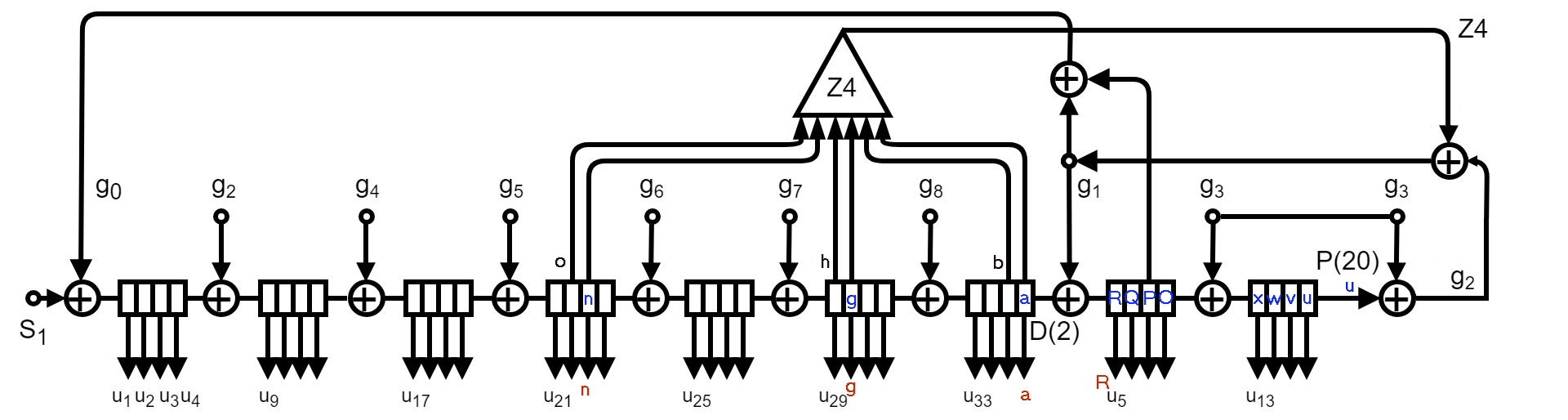}
\end{center}
\vskip-3pt
\vskip-6pt
\caption{
A detailed explanation of what happens with the pre-conditions
specified in Thm. \ref{ThmKT1Cycle9InsideSection9}.3. 
Connections of the inputs of $Z4$ are decided at a later $FE$ solving stage.
Most of the inputs $g_i$ coming from other parts of the cipher get eliminated.
}
\label{FigComplicationUnitSKSTotalOrderT-310CaseCycle9P20LZS991}
\vskip-6pt
\end{figure}
\vskip-8pt

This result can be seen as a small modification of a linear invariant
with an interesting partitioning of products in $W$:
half of them $an+gn$ are connected to inputs, and half are connected to outputs with $bo+ho$.
Syntactically, under very specific quite strong assumptions (above + KT1) it becomes easy to just check
that a specific invariant situation specified here just works.
However, aren't we studying a set of constraints which is contradictory? If so we would be able to prove potentially anything.
A number of further ``sanity'' checks are needed.
Given the fact that the KT1 spec is very complex,
do any LZS with all these properties exist at all? 
These questions are very hard to answer syntactically (or mathematically),
we need to look at specific examples (cf. LZS 991 below).
Moreover, on the computational side, can we actually find KT1 keys,
which satisfy all the constraints of Thm. \ref{ThmKT1Cycle9InsideSection9}.3. 
above,
plus some 15 constraints which define KT1 keys \cite{T-310An80,MasterPaperT310}?
Then also, can they be such that no simpler and trivial/linear invariants exists?
Overall this may seem an incredibly difficult problem.
In fact in practice it was not so difficult. We did not even need to use a SAT solver to solve this.
One can also use the open source software tool keygen.py described in Appendix I.11 of \cite{MasterPaperT310}
which allows to search for KT1 keys with specific constraints on $D()$ and $P()$.
We found plenty of solutions(!).
For example this one:

\begin{verbatim}
991: P=34,4,33,8,31,28,5,27,9,26,32,19,24,18,21,12,17,
25,20,16,36,35,30,29,23,22,7 D=0,36,4,8,12,20,24,28,32
\end{verbatim}

the FE is
\vskip-6pt
\vskip-6pt
$$
W+a+gn+ho+an+bo=0
$$
\vskip-1pt

which equation has a unique solution:
%
$Z=a+ae+ce+bf+df$.

\vskip-8pt
\vskip-8pt
\subsection{A Balanced Invariant}
\label{SimpleInvP20Cycle9Balanced}
\vskip-4pt

It is easy to see that our last invariant
$
{\cal P}=an+gn+u+v+w+x+O+P+Q+R
$
is balanced, i.e. we do not obtain an internal bias inside the cipher.
We are not yet able to apply the ciphertext-only attack
of Section \ref{CiphertextOnlyAttacks}.

\vskip-8pt
\vskip-8pt
\subsection{A Biased Invariant which is Flawed}
\label{SimpleInvP20Cycle9BiasedFEreducedto0}
\vskip-4pt

Interestingly, by the same method we found another KT1 key,
which accidentally has another more complex invariant, still quadratic, which is biased.
Yet ${\cal P}$ is not irreducible and this invariant is far from being OK,
cf. Appendix \ref{SimpleInvP20Cycle9BiasedFEreducedto0App}.

\vskip-8pt
\vskip-8pt
\subsection{A Better Biased Invariant}
\label{SimpleInvP20Cycle9BiasedFEHomQuad}
\vskip-4pt

We need to construct yet another invariant, such that ${\cal P}$ is irreducible and biased.
Again we worked by paper and pencil aiming at an invariant with fewer constraints,
which unlike some earlier examples
would not contradict Thm. \ref{EveryKT1CycleThm}.

\vskip-4pt
\vskip-4pt
$$
\begin{cases}
D(9)=4\cdot 6
\cr
D(6)=4\cdot 8
\cr
(g+o)Y(P[8,10,11,12])\equiv mg+mo 
\cr
Y\not\equiv 0
\end{cases}
$$
\vskip-3pt

\noindent
Here inside our $FE$ is already specified and the notation means that $g$ and $o$ must be inputs with numbers in $\{8,10,11,12\}$
of $Z2=Y$ because $7$ and $9$ are excluded by
KT1 rules \cite{MasterPaperT310}, and $m$ can be some other input of $Z2$.
Moreover $F$ is eliminated (similar as before, due to the choice of bits involved in our invariant).
Moreover we have used the peculiar property that $P(6)=D(8)$ in KT1 spec cf. \cite{MasterPaperT310}
which is another place where we might think that the KT1 spec is actually helping the attacker
(same method could be used with $D(7)$ and $P(13)$ which also come from KT1 rules!).  
Then we have done the usual analysis of cycles on small monomials and combined 4 cycles, in order
to generate a highly plausible invariant:

\vskip-4pt
\vskip-4pt
\vskip-4pt
$$
{\cal P}=eg+fh+eo+fp+gm+hn+mo+np
$$
\vskip-4pt
%


and with the same software as before we found a full KT1 key 

\vskip-5pt
\vskip-5pt
\begin{verbatim}
551: P=17,4,33,12,10,8,5,11,9,30,22,24,20,2,21,34,1,25,
13,28,14,16,36,29,32,23,27 D=0,12,4,36,16,32,20,8,24
\end{verbatim}
\vskip-3pt

which gives the exact same $FE$ which we have anticipated:

\vskip-4pt
\vskip-4pt
$$
Yg+Yo+gm+mo
$$
\vskip-4pt

and one solution is $Z=1+d+e+f+de+cde+def$.
Moreover, ${\cal P}$ is irreducible
and we have checked that no linear approximations true with probability 1 exist for LZS 551.
Therefore this example is not degenerate in the sense
of Appendix \ref{SimpleInvP20Cycle9BiasedFEreducedto0App}
and cannot be obtained with multiple linear approximations.

Furthermore, we need to evaluate the bias of this invariant.
A quick calculation 
shows that we get a partition in two sets of unequal sizes.
Furthermore, we checked that already  with $N=3$ we obtain strong biases
on $N$-tuples of variables which can be used in our 
higher-order correlation attack 
cf. Section \ref{CiphertextOnlyAttacks}.
For example when ${\cal P}=0$
there are 40 events with $efg=1$
and 16 events with $ef(g+1)=1$.

\vskip-6pt
\vskip-6pt
\subsection{How to Make GLC Work?}
\vskip-4pt

It is easy to see that even when $deg({\cal P})=2$
the space of GLC attacks cannot be explored systematically:
we have $2^{{36 \choose 2}}\approx 2^{600}$ possible invariants ${\cal P}$ to try,
and it is not true that ${\cal P}$ must be symmetric.
Both this paper and another recent paper on this topic \cite{TodoNL18}
have (temporarily) studied quadratic invariants mainly.
We also found countless more complex invariants.
It seems that the ONLY way we are aware of 
to at least partly explore this space, is through solving our FE equation.
Moreover, because as it seems, there are plenty of solutions of all sorts and shapes,
we need additional ad-hoc rules and heuristics. 
The main guiding principle we adopted here is to search for invariants which eliminate
specific bits which strongly depend on the key and other parts of the cipher,
e.g. in our specific targeted elimination strategy
in Section \ref{ExampleF0F1andL0L01withZ1andZ4Spaces}.

\vskip-6pt
\vskip-6pt
\subsection{How Many KT1 Keys Can Be Broken?}
\label{SimpleInvP20Cycle9challenge}
\vskip-4pt

We have eventually found several ways of generating invariants which are compatible with KT1 keys.
Our last two invariants with $P(20)$ in Section \ref{SimpleInvP20Cycle9KT1} and in Section \ref{SimpleInvP20Cycle9BiasedFEreducedto0}
work with a cycle of (maximal) length 9, same as with historical keys
and consistent with Thm. \ref{EveryKT1CycleThm}
and with ALL the very complicated 15+ KT1 rules,
cf. \cite{T-310An80,MasterPaperT310}.
We expect that with similar invariants one can break further keys of type KT1
in a ciphertext-only attack scenario following Section \ref{CiphertextOnlyAttacks}.
A recent evaluation in Section 5.4. of \cite{LCKT1ucry18} shows that
the space of KT1 keys has approximately
$2^{83}$ 
elements.
Then it is easy to see that the proportion of KT1 keys
compliant with the exact constraints imposed inside Thm. \ref{ThmKT1Cycle9InsideSection9}.3. 
is about $2^{-23}$.
Then approximately $2^{60}$ keys will
have an invariant attack
specified in Thm. \ref{ThmKT1Cycle9InsideSection9}.3. 
In Section \ref{SimpleInvP20Cycle9BiasedFEHomQuad} there are less constraints
and the number of LZS which can be attacked will be very roughly about
$2^{83}\cdot 9^{-2}\cdot 36^{-2}\approx 2^{65}$.

\vskip-6pt
\vskip-6pt
\subsection{Open Problems}
\vskip-6pt
\label{SimpleInvP20Cycle9BiasedIOpenHO}

In Section \ref{MultipleInterestectElimF} we found two very good invariant attacks which are compatible
with a higher-order correlation attack in Section \ref{CiphertextOnlyAttacks},
however they are not compatible with Thm. \ref{EveryKT1CycleThm}.
We also found invariants which are compatible with Thm. \ref{EveryKT1CycleThm}
but not with our correlation attack.
Eventually both are achieved in Section \ref{SimpleInvP20Cycle9BiasedFEHomQuad} for some
$2^{65}$ LZS.
We conjecture that the actual number of KT1 keys broken by the exact attacks of this paper
should be closer to $2^{80}$ than to $2^{65}$ (out of $2^{83}$ total).
The attacker needs to specify MORE classes of invariants such as in Section \ref{SimpleInvP20Cycle9BiasedFEHomQuad}
and with ${\cal P}$ being 
irreducible and biased
(cf. Appendix \ref{SimpleInvP20Cycle9BiasedFEreducedto0App}),
and add the resulting probabilities.
This sort of result just by the rules of probability would imply a fair chance but no guarantee
of also breaking some historical LZS keys.

\newpage
\vskip-6pt
\vskip-6pt
\section{Conclusion}
\vskip-3pt

In this paper we study cryptanalysis with non-linear polynomial invariants.
We show how a specific structure and internal wiring of
more or less any block cipher, starting from round ANFs,
can be translated into a relatively simple ``Fundamental Equation'' (FE),
which can be used to study 
which specific non-linear invariants may exist (or not) for this cipher.
Stronger invariants can now be defined and characterized algebraically,
${\cal P}$ must be such that of FE reduces to zero,
cf. Section \ref{FEReductionToZeroExample2b}.
In current research in Partitioning Cryptanalysis (PC) \cite{HarpMassThm}
there are some impossibility results \cite{BeiCantResNL,FiliolNotVuln}
but extremely few possibility results \cite{invglc}. 
Partitioning properties are extremely hard to find.
Polynomial invariants are way more intelligible.
Our main contribution is to show that the attacker does not need
to randomly search for an interesting invariant ${\cal P}$
and a vulnerable non-linear component $Z$.
Specific polynomial invariants ${\cal P}$ and
weak Boolean functions which work together
can now be {\bf determined} -- by solving our $FE$ equation(s).
Our approach is constructive, completely general and can be applied to almost any block cipher:
we first write the $FE$ 
then based on ad-hoc 
heuristics we determine a space of polynomials
with a reduced dimension for ${\cal P}$,
we substitute variables inside the $FE$(s),
and we attempt to solve our $FE$(s).
We have constructed numerous 
concrete examples of non-trivial non-linear invariants 
which propagate for any number of rounds, and for any key and IV.

We anticipate that the success rate of this approach will be very different
for different families of ciphers. 
If just one round function is very complex and uses many key bits,
with too many constraints to satisfy simultaneously,
our approach is likely to fail,
which could be used in security proofs, cf. Section \ref{SecSecurityProofs}.
Or solving FE will become computationally difficult,
which would not mean that the cipher is secure,
rather it could be broken tomorrow at a cost of a larger one-time pre-computation.


\vskip-6pt
\vskip-6pt
\subsection{Cryptanalytic Applications: Biased Partitioning} 
\vskip-2pt

We show that some invariants are trivial and result from Linear Cryptanalysis
and the polynomial ${\cal P}$ is a highly symmetric product of linear factors.
In many other cases, ${\cal P}$ is however not symmetric and irreducible.
Then typically we obtain a partition
into two sets with similar yet {\bf unequal sizes}.
Such invariants 
introduce a {\bf permanent and pervasive bias} inside the cipher
which is not degraded
with iteration of the cipher.
This immediately leads to ciphertext-only higher-order correlation attacks with key recovery
such that their complexity does NOT depend on the number of rounds, cf. Section \ref{CiphertextOnlyAttacks}.


%

\vskip-6pt
\vskip-6pt
\subsection{Cryptanalysis of Stronger KT1 Keys}
\vskip-3pt

What is also quite incredible is that our attack works, at all,
for at least one real-life block cipher. 
With our methodology we are able to generate lots of invariants of all sorts of shapes,
more or less on demand, for example invariants such that two remote parts of the cipher interact,
in such a way that other very complex parts which connect them
are eliminated, and moreover key bits are also eliminated, cf. Section \ref{ExampleF0F1andL0L01withZ1andZ4Spooky}
and \ref{ExampleF0F1andL0L01withZ1Z2Z4}.
Now, can we also construct invariants which are compatible with
Thm. \ref{EveryKT1CycleThm}
which characterizes a class of 
stronger KT1 keys which were specified by the designers?
Yes, and as a proof of concept in Section \ref{MysteryPtySection}
we present three 
methods to construct KT1 keys
with an invariant ${\cal P}$ which propagates for any number of rounds and any key and $IV$.

\vskip-5pt
\vskip-5pt
\subsection{On Vulnerable Boolean Functions}
\label{ConclHighDegPower1}
\vskip-3pt

In most of our proof of concept examples the Boolean function is very special
or quite trivial and has generally a lower degree than expected.
This is probably not a problem and is due to the fact that we impose yet very few constraints
and our examples have been chosen for elegance and simplicity.
An important point is that at this moment we have {\bf artificially} imposed that many inputs of these
Boolean functions are not used in our $FE$ or could be eliminated totally. 
This inevitably leads to quite peculiar Boolean functions.
This problem will probably 
go away as soon as more complex variants using more bits as inputs are constructed.
Moreover, in general, when the degree of ${\cal P}$ increases we expect to find many more/stronger invariants ${\cal P}$
which work for a larger proportion of the space of all Boolean functions on 6 variables.
An open problem is to find examples which would work in the worst case
(for ANY Boolean function and when the Fundamental Equation is reduced to 0)
which would not be degenerate in the sense of Appendix \ref{SimpleInvP20Cycle9BiasedFEreducedto0App}.

\medskip
\noindent
{\bf Latest News - Added in 2019:}
A first attack and proof of concept showing that when the degree of ${\cal P}$
increases one can attack indeed more or less arbitrarily strong Boolean functions
was found and published in December 2018, see \cite{BackdAnn}.

\vskip-5pt
\vskip-5pt
\subsection{On Success Rate of Our Attacks}
\label{ConclHighDegPower2}
\vskip-3pt

We conjecture that for any Boolean function there exists one or several
low degree invariants $P$ such that T-310 is broken by our ciphertext-only attack,
and that the percentage of LZS which are vulnerable to our attack
should
increase
when the degree of ${\cal P}$ grows.
We also expect that the percentage of vulnerable Boolean functions increases.
In Section \ref{SimpleInvP20Cycle9challenge} we have estimated that at least
$2^{65}$ out of $2^{83}$ of all possible KT1 keys
will already be broken by the simple invariants which we already found.
Many more can yet be broken by the same attack
and there are some hopes 
that some actual historical KT1 keys could also be broken.
We have just started to explore an incredibly rich space of new attacks.


\newpage

\appendix
\vskip-6pt
\vskip-6pt
\section{A Degenerate KT1 Invariant with FE Reduced to Zero}
\label{SimpleInvP20Cycle9BiasedFEreducedto0App}
\vskip-4pt

Not all non-linear invariants are good.
The following quite special and degenerate case was found accidentally during our search for KT1 keys
in Section \ref{SimpleInvP20Cycle9KT1}:

\begin{verbatim}
881: P=4,20,33,8,1,28,5,19,9,32,11,17,24,13,21,18,15,
25,12,16,35,22,23,29,36,30,34 D=0,36,4,8,12,20,24,28,32
\end{verbatim}

Here the invariant ${\cal P}$
is a homogenous polynomial of degree 2 with $169=13^2$ terms and 26 variables
which is highly symmetric and not at all irreducible:

\vskip-6pt
\vskip-6pt
$$
(n+b+p+r+t+v+x+z+N+P+R+T+V)
(a+m+o+q+s+u+w+y+M+O+Q+S+U)
$$
\vskip-4pt


Furthermore it is possible to verify that
if we call ${\cal A}$ the first sum,
and if we call ${\cal B}$ the second sum,
${\cal A}{\cal B}$ is a non-linear invariant
and also
${\cal A}+{\cal B}$ is a linear invariant for 1 round and the exact same cipher setup LZS 881.
The linear invariant produces a partition with two sets of equal sizes.
For the non-linear invariant the sizes are not
equal and the probability that 
${\cal P}={\cal A}{\cal B}=0$ is exactly 3/4.
Moreover, it is easy to see that this invariant works also for the original Boolean function
(actually the $FE$ is reduced to 0 here) and also for any other $Z$.
This example remains however very special.
Both ${\cal A}$ and ${\cal B}$ one at a time are 2-round
invariants with ${\cal A}\mapsto {\cal B}$ after 1 round.
Then after one round ${\cal A}\cdot {\cal B}$
becomes
${\cal B}\cdot {\cal A}$ which is the same, hence a quadratic invariant for 1 round exists.
This is not a very good invariant: the 3/4 bias obtained here
is something which we would already be constructed by combining the two linear approximations.
A detailed study shows that up to 13 linear properties can be made to hold simultaneously in T-310,
cf. Section 21.16 in \cite{MasterPaperT310} and up to 10 can be obtained for KT1 keys.
This leads to plenty of non-linear invariants which
can be constructed from linear invariants (only if not using $F,S1,S2$ bits)
which can be multiplied (or added) together to form a variety of trivial invariants.
Most likely this can also be done with linear invariants using $F,S1,S2$ bits
in such a way, which would not trivial,
that these bits will be eliminated totally, and in the non-linear version only.
Then the {\bf only} invariants which do not depend on key and IV bits would be non-linear invariants.
However overall this is not the sort of invariants we aim at constructing in this paper.
As shown in this paper, the most interesting invariants for a cryptanalyst will be those
where ${\cal P}$ is irreducible and yet the partitioning gives two sets of unequal sizes.

{\bf Remark:} It is easy to construct degenerate examples also when ${\cal P}$ is irreducible:
for example through linear combinations by adding two or more products such as above.
Typically in such cases the partitioning also gives two sets of unequal sizes,
which however do not give anything compared to the Linear Cryptanalysis.


\newpage

\vskip-6pt
\vskip-6pt
\section{Frequently Asked Questions, Q/A, Discussion/Rebuttal}
\label{RebuttalQuestions}
\vskip-4pt

This Appendix section can be a seen
as an alternative post-scriptum discussion of this paper added after we have
received some feedback
from peer crypto researchers at 
Eurocrypt 2019.
In order to understand better this Q/A section,
it is important to see that many IACR conferences such as Eurocrypt 2019
have now a formal rebuttal phase, 
and we were given an opportunity to answer (space permitting)
to the questions raised by the reviewers.
The comments received show among other,
that maybe we need to explain certain things better,
or that our research is highly innovative and if
the reviewers did not get some points, many other people will not either,
or that the PC has simply decided that this sort of research
cannot or should not at all\footnote{
All the referees were very clear on this point,
this paper should not be presented at Eurocrypt,
based on the ideas that T-310 is old or ``not relevant''.
In reality this paper shows that the research community have never ever
yet started studying the topic of block ciphers properly
and ignored the very existence of an incredibly
vast space of powerful and general attacks,
a question reaching far beyond any block cipher in particular.
}
be published at Eurocrypt
for some specific 
reasons.

This paper is in fact almost entirely identical to the paper we submitted to Eurocrypt 2019
and all changes made were made in order to improve clarity.
Below we respond to the comments and provide short answer to these questions in a straightforward direct way,
hoping that our work can be better understood.
We have {\bf not} added any new content here\footnote{
This paper is past-looking,
a record of our knowledge in October 2018, and we made a deliberate choice not to add any new research here,
except to note that our bold conjectures in our conclusion Section \ref{ConclHighDegPower1}
are now confirmed in \cite{BackdAnn}.}
except for the sake of clarifying the exact questions raised by the reviewers.
The exact rebuttal text submitted to Eurocrypt is included below.
Not a word was added or removed.
The language is quite laconic and direct due to severe space limitations.



{
\begin{verbatim}
"a full definition": a FULL description of the block cipher inside T-310
is given TWICE: formulae p. 7, and Fig. 1+2+P/D

"known security recommendations"?? For this we cite 2017/440.

"I do not see how this implies attacks on a given instance":
works because two polynomials are equal. Invariant propagates
for any nb. of rounds and any key, strong pty.

"results on relevant ciphers"? Huge but indirect.
A new extremely rich algebraic method to analyse the security
of ANY block cipher:
FE method + RowEchelon elim sec.7.3.
We do not claim it breaks 3DES... yet please check!
How many attacks ever seen work for any number of rounds?
100% of crypto research have developed ignoring ABCs with
misplaced priorities: "black box" approximations for N rounds?
"more complex" and stronger attacks may exist, 1R=> game is over.
New white-box model, attacks are derived mathematically.
Stop ignoring complex attacks.
Huge space => any positive working examples valuable.

"variable names are taken from some source code"???
if so, they would be long, in fact 1 letter, cannot be shorter.

Real-life Z1 to Z4 are fixed. Yet researchers must explain HOW Z is chosen!
We show that super-weak choices exist! No idea WHEN exactly this old cipher
is secure => no reason to have an informed opinion on later block ciphers
e.g. DES. Design criteria must be revised: many pties of Z do NOT matter
for larger N, our work is relevant for ANY number of rounds.

"authors first pick an invariant and then see for which function Z1 to Z4
this is actually an invariant" NOT like this! We do sth fundamentally stronger.
Our invariants eliminate variables, and quite many variables. When we use a
variable say M, our invariant ELIMINATES tens of variables needed to compute M!!
At the end an attacks may exist, FE yielding the answer, but P and the FE
eliminate plenty of vars. Cf. Sec 7.2 with vars from 2 opposite sides,
half depend on absolutely EVERYTHING inside.
Yet these bits get eliminated.

"relevant ciphers"???
THE most important government block cipher for the whole of the Cold War
is not relevant for a crypto conference? Father and early cousin
of all block ciphers ever seen? Used by a totalitarian regime?
The research community have never found a complex sophisticated attack
on block ciphers, only simple ones s.t. space is small [64 bits/small HW].
No idea about even how a complex attack on a complex block cipher
could even look like.
Also almost never studied ciphers with low data extraction rate.
Scientific challenge: 1000S of of rounds => most academic attacks do not work,
ours may work cf. Section 8 - ciphertext-only.

"Luckily, cryptographic research has progressed since the 1980's", maybe...
Or proper research on the security of block ciphers have never yet started!
Bootstrapping problem: no good examples were known.
We still don't know how to select Z for T-310 and never yet studied
99.99999% of attacks where this would matter.
\end{verbatim}
}

\end{document}